%% file: RicciTimeInLT.tex
\documentclass[12pt,A4]{article}

\usepackage[dvips]{graphicx}
\usepackage{amsmath}
\usepackage{amsfonts}
\usepackage{amssymb}
\usepackage{longtable}
\usepackage{lscape}
\usepackage{subfig}
\usepackage{color}
\usepackage{pst-plot}
\usepackage{pstricks}
\usepackage{multido}
\usepackage{fp}
\usepackage{ifthen}

\allowdisplaybreaks[4]     % allows page breaks in the middle of an equation group.

\newcommand{\er}[1]{(\ref{#1})}          % Equation reference with brackets
\newcommand{\nn}{\nonumber}
\renewcommand{\d}{{\rm d}}     % Normally  \d{o}  gives an underdot.
\newcommand{\td}[2]{\frac{{\rm d} {#1}}{{\rm d} {#2}}}
\newcommand{\tdil}[2]{{\rm d} {#1} / {\rm d} {#2}}

\newcommand{\pdil}[2]{\partial {#1} / \partial {#2}}

\newcommand{\showlabel}[1]{
  \label{#1}
}
 \newcommand{\pb}[2]{                       % Simplified tailored parbox
    \parbox[t]{#1}{
       \raggedright
       \setlength{\parskip}{1.2ex}          % Set to same as document default
       #2
    }
 }
\renewcommand{\L}{Lema\^{\i}tre}
\newcommand{\LT}{\L-Tolman}

\title{Ricci Time in the \LT\ Model and the Block Universe}
\author{
  Yasser Elmahalawy\thanks{ELMYAS001@myuct.ac.za,yasser@aims.ac.za}, 
  Charles Hellaby\thanks{Charles.Hellaby@uct.ac.za},
  and
  George F.R. Ellis\thanks{George.Ellis@uct.ac.za}
  \\
  {\small Department of Mathematics and Applied Mathematics,} \\
  {\small University of Cape Town, Rondebosch, 7701, South Africa}
}
\date{}

\begin{document}
\sffamily

\maketitle

\begin{abstract}
It is common to think of our universe according to the ``block universe" concept, which says that spacetime consists of many ``stacked" 3-surfaces, labelled by some kind of proper time, $\tau$.  Standard ideas do not distinguish past and future, but Ellis' ``evolving block universe" tries to make a fundamental distinction.  One proposal for this proper time is the proper time measured along the timelike Ricci eigenlines, starting from the big bang.  This work investigates the shape of the ``Ricci time" surfaces relative to the the null surfaces.  We use the \LT\ metric as our inhomogeneous spacetime model, and we find the necessary and sufficient conditions for these $\{\tau$= constant$\}$ surfaces, $S(\tau)$, to be spacelike or timelike.  Furthermore, we look at the effect of strong gravity domains by determining the location of timelike $S$ regions relative to apparent horizons.  We find that constant Ricci time surfaces are always spacelike near the big bang, while at late times (near the crunch or the extreme far future), they are only timelike under special circumstances.  At intermediate times, timelike $S$ regions are common unless the variation of the bang time is restricted.  The regions where these surfaces become timelike are often adjacent to apparent horizons, but always outside them, and in particular timelike $S$ regions do not occur inside black holes.
\end{abstract}
% {\large [[[* The equation labels are suppressed by commenting out line 37 in the file {\tt RicciTimeLT.tex} *]]]}

\input{Section1}
\input{Section2}
\input{Section3}
\input{Section4}
\input{Section5}

%\appendix

%\input{AppendixA}
%\input{AppendixB}

\bibliography{Bibliography}{}
% Bibliography style options are (see http://sites.stat.psu.edu/~surajit/present/bib.htm):
% abbrv = plain = numbered, in surname alphabetical order,
% acm,
% alpha,
% apalike, 
% ieeetr = numbered, in order cited in text,
% plain = abbrv
% siam = 
% unsrt = 
\bibliographystyle{plain}

\end{document}

%% file: Section1.tex
%%%%%%%%%%%%%%%%%%%%%%%%%%%%%%%%%%%%%%%%%%%%%%%%%%%%%%%%%%%%%%%%%%%%%%%%%%%%%%%%%%%%%%%%%%%%%%%%%%%%
\section{Introduction}
\showlabel{Section1}

There are numerous theories that discuss the nature of time and the way to measure time. Rovelli \cite{Rovelli:2009ee} and Barbour \cite{barbour99} assumed that time does not exist and does not ``roll on'', according to the idea of standard block universe. Davies \cite{Davies_2012} proposed that time is real but flow is not. Price \cite{price96} and Barbour \cite{barbour99} argue that a block universe representation denies that time passes and embodies the idea that the flow of time is an illusion. Ellis \cite{Ellis:2012ay} examined the various views and argued that time exists and its flow is real, calling this the evolving block universe (EBU).
%The idea of a block universe \cite{Butterfield:2001fh, FQXi_2014, Mellor98} was proposed by Price \cite{price96} and Barbour %\cite{barbour99}. This representation denies that time passes and embodies the idea that the flow of time is an illusion %\cite{Butterfield:2001fh} and time does not really ``roll on". An alternative is the idea of an evolving block universe, enabling one  to envisage the flow of time and give a clear picture of a universe where things change \cite{Ellis_2002}. 
This idea allows us to contemplate the different natures of the past and future, and differentiate between them in our spacetime representation. Moreover, it shows how to indicate the present ``now" in the spacetime diagram. \\

%We briefly summarize the evolution of this theory, based on either classical mechanics \cite{Feynman63} or quantum mechanics %\cite{Feynman651}, and supported by aspects of General Relativity \cite{Hawking_Ellis_1975, ellis2000}. The 
This paper will not repeat these discussions: the main issue we address is the concept proposed in \cite{Ellis:2012ay} of how 
to characterize preferred time surfaces in a curved spacetime filled with a non-zero energy density (for example the Cosmic Microwave Background Radiation, or CMB). This proposal can be investigated independently of its relation to the EBU idea, as it is an interesting geometrical question in its own right as to how the preferred spatial sections mentioned in \cite{Ellis:2012ay} evolve.

According to the EBU, the universe consists of a sequence of curved surfaces, labelled by proper time $\tau$, which is measured along a family of chosen timelines $x^{i}(\upsilon)$  from the start of the universe to the present day. Ellis \cite{Ellis:2012ay} argued that the present time surface at proper time $\tau=\tau_0$ is the surface of \{$\tau=\tau_0$\} defined in this way, which can be measured by integration along a family of fundamental world lines; as time passes, $\tau_0$ continually grows, so the future boundary of spacetime is always extending (hence the name EBU). The fundamental world lines used in this definition are chosen to be the Ricci eigenlines, which represent the average motion of matter, and will be well defined in all realistic spacetimes, because of the existence of an all-pervading cosmic blackbody radiation in the universe.

This prescription defines a family of surfaces that may be locally either spacelike, null, or timelike. The main idea of this paper is to examine the behavior of the \{$\tau=constant$\} surfaces $S(\tau_{c})$ and find out their nature: are they locally spacelike, timelike, or null? This is an interesting geometric question that can be asked independently of whether one accepts the idea of an evolving block universe or not. One might expect that these surfaces will be spacelike except in the case of very intense gravitational fields. We will show, by investigating these surfaces in a number of \LT\ models, that this is not necessarily the case. 
 
The \LT\ (LT) metric is an exact solution of the Einstein field equations (EFEs), with a dust equation of state, suitable for studying the possibility of an inhomgeneous universe, or inhomogeneous structures within it. Like the Friedmann-\L-Robertson-Walker (FLRW) model, it is spherically symmetric with comoving matter, but \LT\ models are generally inhomogeneous in the radial direction. It was first proposed by Lema\^{\i}tre \cite{Lemaitre_1933} in 1933 and then by Tolman \cite{Tolman_1934} in 1934; a few years later, in 1947, Bondi \cite{Bondi_1947} investigated this model again.  It has been widely used to model large and small scale structures, and to study gravitational collapse, cosmological observations, and exact nonlinear features of the EFEs \cite{Hellaby_2006m, Hellaby_Alfedeel_2009, Bolejko_Krasinski_Hellaby_2005, Alnes_Amarzguioui_Gron_2006, Sussman_2007}.  Interestingly, it can be used to describe a homogeneous cosmology in one limit, and non-vacuum black holes in another limit  \cite{Hellaby_1999, Krasinski_Hellaby_2004a}.  See \cite{andrzej97,Plebanski2006} for a summary of work on the \LT\ model, and \cite{Hellaby:2009vz} for a well illustrated outline.

%% file: Section2.tex
%%%%%%%%%%%%%%%%%%%%%%%%%%%%%%%%%%%%%%%%%%%%%%%%%%%%%%%%%%%%%%%%%%%%%%%%%%%%%%%%%%%%%%%%%%%%%%%%%%%%
\section{The \LT\ Model and its Evolution}
\showlabel{Section2}

%%%%%%%%%%%%%%%%%%%%%%%%%%%%%%%%%%%%%%%%%%%%%%%%%%%%%%%%%%%%%%%%%%%%%%%%%%%%%%%%%%%%%%%%%%%%%%%%%%%%
\subsection{\LT\ Model}
\showlabel{lt}
The \LT\ (LT) metric is written in spherically symmetric, synchronous coordinates,
\begin{align}
\showlabel{lt1}
\d s^{2} & =  - \d t^{2} + \frac{(R')^{2}}{1+f} \, \d r^{2} + R^{2} (\d \theta^{2} + \sin^{2}\theta \, \d \phi^2),
\end{align}
where $R$ is the areal radius, $f(r)$ is a free function determining the local geometry, and $R^{'} = \pdil{R}{r}$. The matter is a pressure-free perfect fluid \cite{Ellis_1967}, comoving with the coordinates, and the energy momentum tensor is defined by
\begin{align}
\showlabel{lt2}
T^{ab} & =  \rho u^{a} u^{b},
\end{align}
where $\rho$ is the mass-energy density, and $u^{a}$ is the fluid's four velocity,
\begin{align}
\showlabel{lt3}
u^{a} & =  \delta^{a}_{t}.
\end{align}
The solution of the EFEs gives the evolution equation
\begin{align}
\showlabel{lt4}
\dot{R}^{2} & =  \frac{2M(r)}{R}+f(r) + \frac{\Lambda R^{2}}{3},
\end{align}
where $\dot{R} = \pdil{R}{t}$, the second free function $M(r)$ is the gravitational mass within the comoving shell of radius $r$, and $f(r)$ has an additional interpretation as $f = 2E(r)$ which is twice the local energy per unit mass of the dust particles. The EFEs also give the density as
\begin{align}
\showlabel{lt5}
\kappa \rho & =  \frac{2M'}{R^{2}R'}.
\end{align}
The Kretschmann scalar \cite{Hellaby:2009vz} is an invariant measure of spacetime curvature which is given by
\begin{align}
\showlabel{lt5a}
\mathcal{K} & =  R_{abcd} R^{abcd} = \frac{48M^{2}}{R^{6}} + \frac{32 M M'}{R^{5} R'} +\frac{12 (M')^{2}}{R^{4} (R')^{2}};
\end{align}
$\mathcal{K}$ only diverges where $R$ or $R'$ are zero, while $M$ and $M'$ are not.  Similarly, $\rho$ only diverges where $M'/R'$ is zero.
The solutions of \er{lt4}, when the cosmological constant $\Lambda$ is zero, are of 3 types, depending on the value of $f$ (or more correctly, the value of $f/M^{3/2}$),
\begin{align}
\showlabel{lt6}
& \mbox{Hyperbolic, $f>0$}:~~ &
R & =  \frac{M}{f}(\cosh \eta -1),~~ & (\sinh \eta - \eta) & = \frac{f^{3/2}(t-a)}{M}; \\
\showlabel{lt7}
& \mbox{Parabolic, $f=0$}:~~ &
R & = M \left( \frac{\eta^{2}}{2} \right),~~ & \left( \frac{\eta^{3}}{6} \right) & = \frac{(t-a)}{M}; \\
\showlabel{lt8}
& \mbox{Elliptic, $f<0$}:~~ &
R & = \frac{M}{(-f)}(1 - \cos \eta),~~ & (\eta - \sin \eta) & = \frac{(-f)^{3/2}(t-a)}{M};
\end{align}
where the last free function $a(r)$ is the local time of the big bang, i.e. the time on each worldline at which $R = 0$.  The bang time does not have to be the same for each worldline; usually outer shells of space and matter ``explode" off the initial singularity at earlier times.

The parameter $\eta$ can be thought of as giving the stage of evolution, such as ``early", ``intermediate", and ``late".  The 3 types of solution are not mutually exclusive, and it is possible to have a hyperbolic region outside and elliptic region, with a parabolic locus at the boundary between them; this models gravitational collapse in an expanding universe \cite{Hellaby_Krasinski_2001, Krasinski_Hellaby_2004}.  Though extended parabolic regions are possible, they are all asymptotically homogeneous, consequently they will not be investigated explicitly here.  The 3 arbitrary functions of this model allow a freedom to rescale the $r$ coordinate plus two physical freedoms, such as $a(M)$ and $f(M)$.  However, as noted in \cite{Hellaby:2009vz}, each function contributes to the physical inhomogeneity, and specific ``coordinate choices" restrict physical possibilities. For example, if $M = r^3$ is specified, then vacuum regions $M' = 0$ are excluded; similarly for the other two.

A scale length and time, characteristic for each worldline, are defined as
\begin{align}
\showlabel{lt12}
\tilde{R}(r) = \frac{M}{|f|}, ~~~~~~\mbox{and}~~~~~~
\tilde{T}(r) = \frac{M}{|f|^{3/2}},
\end{align}
respectively. Note that the lifetime from the bang to crunch in elliptic models is $2 \pi \tilde{T}$.

%%%%%%%%%%%%%%%%%%%%%%%%%%%%%%%%%%%%%%%%%%%%%%%%%%%%%%%%%%%%%%%%%%%%%%%%%%%%%%%%%%%%%%%%%%%%%%%%%%%%
\subsection{Singularities}
Apart from the big bang and big crunch singularities, which are non-simultaneous versions of their FLRW counterparts, LT models may also suffer from shell crossing singularities, as well as ``shell focussing" singularities.  The big bang occurs on each worldline where $t=a$, causing $R(a, r) = 0$ for all $r$.  The big crunch occurs in elliptic models where $t=a + 2 \pi \tilde{T}$, also causing $R = 0$ on each worldline.  The bang and crunch 3-surfaces are space-like \cite{Hellaby_1985} as will be seen in section \ref{Section3}, and have diverging density \er{lt5} and curvature \er{lt5a}.

Shell crossings occur due to the collision between an inner and an adjacent outer shell of constant $r$, meaning $R'=0$ locally; consequently density and curvature diverge.  The shell-crossing surfaces are time-like, and they have a different redshift structure from the big bang\cite{Hellaby_Lake_1986}.  Though matter collision may be allowable in a much more sophisticated matter description, shell crossings are physically unacceptable in LT models.  The conditions on the LT arbitrary function that ensure no shell crossings are avoided were given in \cite{Hellaby_Lake_1986,Krasinski_Hellaby_2004a}.  See table \ref{S}. 

Shell focusing singularities can occur on the central worldline (origin), at the moment of the big crunch.  Under certain conditions, many light rays can be emitted from this one point on the crunch and can even travel to infinity, rather than being captured by the crunch \cite{Joshi93, Waugh_Lake_1989, Newman_1986, Eardley_1979, Christodoulou_1984}.  It will not be considered here.

\begin{table}[!ht]
\centering
\begin{tabular}{| p{2cm} | p{2cm} | p{6cm} |}
\hline
$R^{'}$ & $f$ & $M^{'}$, $f^{'}$, $a^{'}$ \\ \hline
$>0$ & $\geq0$ &
  \parbox[t]{6cm}{$M^{'} \geq 0$ \\ 
  $f^{'} \geq 0$ \\ 
  $a^{'} \leq 0$ \\ 
  but not all 3 equalities at once} \\ \cline{2-2} \cline{3-3}
& $<0$ &
  \parbox[t]{6cm}{$M^{'} \geq 0$ \\
  $2 \pi \tilde{T}^{'} + a^{'} \geq 0$ \\
  $a^{'} \leq 0$ \\
  but not all 3 equalities at once
  } \\ \hline
$=0$ & $=0$ &
  \parbox[t]{6cm}{$M^{'} = 0$ \\ 
  $f^{'} = 0$ \\ 
  $a^{'} = 0$
  } \\ \hline
$<0$ & $\geq0$ &
  \parbox[t]{6cm}{$M^{'} \leq 0$ \\ 
  $f^{'} \leq 0$ \\ 
  $a^{'} \geq 0$ \\
  but not all 3 equalities at once} \\ \cline{2-2} \cline{3-3}
& $<0$ &
  \parbox[t]{6cm}{$M^{'} \leq 0$ \\
  $2 \pi \tilde{T}^{'} + a^{'} \leq 0$ \\
  $a^{'} \geq 0$ \\
  but not all 3 equalities at once} \\
\hline \end{tabular}
\caption{Conditions for no shell crossings.} \showlabel{S}
\end{table}

%%%%%%%%%%%%%%%%%%%%%%%%%%%%%%%%%%%%%%%%%%%%%%%%%%%%%%%%%%%%%%%%%%%%%%%%%%%%%%%%%%%%%%%%%%%%%%%%%%%%
\subsection{Origins and Regular Extrema}
\showlabel{Origins}
\showlabel{RegExtr}
An origin occurs where $R(r_{0},t)=0$ for all $t$ on a particular worldline, $r_0$ \cite{Bonnor1985}.  This implies that all time derivatives of $R$ along $r_0$ are also zero.  Obviously, even at the origin, $t$ must vary over a finite or infinite range, and so must the parameter $\eta$.  So it is clear from equations \er{lt6} and \er{lt8} that, $M/|f|^{3/2}$ must remain finite and non-zero in the limit as $r \to r_0$, and $M/|f|$ must go to zero to make $R$ zero.  Therefore, $M \sim |f|^{3/2}$ in the neighbourhood of the origin, and both $M$ and $f$ go to zero there.  But this does not mean that the time evolution is parabolic \cite{Hellaby_1985}.  When doing numerical calculations, we need to avoid calculating zero over zero, so for each choice of arbitrary functions, we need separate expressions for quantities like $M/f^{3/2}$ that have well behaved origin limits.  The combinations of arbitrary functions and derivatives in equations \er{lt9}-\er{lt12} all require this treatment.  In contrast to the big bang, an origin is timelike and non-singular.

In a spatially closed spherical model, such as the $k = +1$ FLRW model, there must be place where the areal radius is maximum, decreasing on either side towards an origin; a closed spherically symmetric model has two origins.  Similarly, in LT models of SKS black holes, vacuum and non-vaccum, there is a minimum of the areal radius at the ``throat" or ``neck".  Regular extrema in LT models occur if $R' = 0$ at one or more particular $r_m$ values without the density or curvature diverging and without a shell crossing forming \cite{Hellaby_Lake_1986}. This means that
\begin{align}
\showlabel{re1}
R'(t,r_{m}) & = 0 \qquad \forall~ t
\end{align}
The conditions for a regular maximum or minimum \cite{Hellaby_Lake_1986} without shell crossings or surface layers are
\begin{align}
\showlabel{re2}
M'(r_{m}) & = f'(r_{m}) = a'(r_{m}) = 0, \qquad f(r_{m}) = -1.
\end{align}
Unlike shell crossings, regular extrema remain at constant $r$, and are not singular.

%%%%%%%%%%%%%%%%%%%%%%%%%%%%%%%%%%%%%%%%%%%%%%%%%%%%%%%%%%%%%%%%%%%%%%%%%%%%%%%%%%%%%%%%%%%%%%%%%%%%
\subsection{Special Cases}
\showlabel{SpCs}
The LT model has several interesting special cases; the first two are of particular interest here.  The dust FLRW model is obtained when
 \begin{align}
\showlabel{Sc1}
f \propto M^{2/3}, \qquad a' = 0.
\end{align}
If $M' = 0$, then we have at least part of the vacuum spherical spacetime.  To get the full Schwarzschild-Kruskal-Szekeres (SKS) spacetime, requires an elliptic region, and a worldline where $f = -1$ and $a'=f'=0$ at which $R' = 0$, $a'$ is maximum, and $f'$ and $R'$ are minimum.  One may easily construct a non-vacuum black hole with the same geometry and topology, if $M'$ is made non-zero and minimum at $f = -1$ \cite{Hellaby_1996, Hellaby_1999}.  The Datt-Kantowski-Sachs metric is a well-behaved limit of the LT model, occuring if $R^{'} = 0$ and $f = -1$ globally \cite{Hellaby_1996}.  The Vaidya metric is the limit in which the matter worldlines are null; after appropriate transformations, the limit of infinite local energy $f \rightarrow \infty$ is taken \cite{Lemos_1992, Hellaby_1994}.

%%%%%%%%%%%%%%%%%%%%%%%%%%%%%%%%%%%%%%%%%%%%%%%%%%%%%%%%%%%%%%%%%%%%%%%%%%%%%%%%%%%%%%%%%%%%%%%%%%%%
\subsection{Parametric evolution of $R'$}
\showlabel{ParEvRr}
Parametric expressions for the evolution of $R'$ \cite{Hellaby_Lake_1986,Hellaby:2009vz} will be very useful for calculating the radial null slopes for both elliptic and hyperbolic cases in section \ref{Section3}. They are derived from \er{lt6} and \er{lt8}:
\begin{align}
\showlabel{lt9}
& \mbox{Elliptic, $f<0$}:~~ &
R' & = \frac{M'}{(-f)} \psi_{1} + \frac{M f'}{f^{2}} \psi_{2} + (-a') \sqrt{(-f)} \psi_{3}, \\
\showlabel{lt9a}
& \mbox{where} &
\psi_{1} & = 2 + \frac{\eta \sin \eta}{(1 - \cos \eta)}, \\
\showlabel{lt9b}
%&& \psi_{2} & = \frac{1}{2}  \frac{4 - 4 \cos \eta - 3 \eta \sin \eta  +  \sin^{2} \eta}{(1 - \cos \eta)}, \\
&& \psi_{2} & = 2 - \frac{\sin \eta (3 \eta - \sin \eta)}{2 (1 - \cos \eta)}, \\
%&& \psi_{2} & = \frac{(5 + \cos \eta)}{2} - \frac{3 \eta \sin \eta}{2 (1 - \cos \eta)} \\
\showlabel{lt9c}
&& \psi_{3} & = \frac{\sin \eta}{1 - \cos \eta}. \\
\showlabel{lt10}
& \mbox{Hyperbolic, $f>0$}:~~ &
R' & = \frac{M'}{f} \psi_{4} + \frac{M f'}{f^{2}} \psi_{5} + (-a') \sqrt{f} \psi_{6}, \\
\showlabel{lt10a}
& \mbox{where} &
\psi_{4} & = \frac{\eta \sinh \eta}{(\cosh \eta - 1)} - 2, \\
\showlabel{lt10b}
%&& \psi_{5} & = - \frac{1}{2}  \frac{4 - 4 \cosh \eta + 3 \eta \sinh \eta  - \sinh^{2} \eta}{\cosh \eta - 1}, \\
&& \psi_{5} & = 2 + \frac{\sinh \eta (\sinh \eta - 3 \eta)}{2 (\cosh \eta - 1)}, \\
%&& \psi_{5} & = \frac{(5 + \cosh \eta)}{2} - \frac{3 \eta \sinh \eta}{2 (\cosh \eta - 1)} \\
\showlabel{lt10c}
&& \psi_{6} & = \frac{\sinh \eta}{\cosh \eta - 1}.
\end{align}

The limiting behaviour of the evolution functions at early and late times, and their values at maximum expansion, will also be useful:
\begin{table}[!ht]
\centering
\begin{tabular}{|l|l|l|l|l|l|l|}
\hline
\multicolumn{4}{|c|}{Elliptic} & \multicolumn{3}{|c|}{Hyperbolic} \\ \hline
& Early & Max Expansion & Late & & Early & Late \\ \hline \hline
$\eta$ & $\to 0$ & $= \pi$ & $\delta = 2\pi - \eta \to 0$ & $\eta$ & $\to 0$ & $\to \infty$ \\ \hline
$\psi_1$ & $\sim \eta^2/6$ & $= 2$ & $\sim 4 \pi/\delta$ & $\psi_4$ & $\sim \eta^2/6$ & $\sim \eta$ \\ \hline
$\psi_2$ & $\sim \eta^4/40$ & $= 2$ & $\sim 6 \pi/\delta$ & $\psi_5$ & $\sim \eta^4/40$ & $\sim e^\eta/4$ \\ \hline
$\psi_3$ & $\sim 2/\eta$ & $= 0$ & $\sim - 2/\delta$ & $\psi_6$ & $\sim 2/\eta$ & $\sim 1 + 2 e^{-\eta}$ \\ \hline
\end{tabular}
\caption{Limiting behaviours of the evolution functions.}
\showlabel{PsiLim}
\end{table}

Finally, the $\eta$ derivatives of the hyperbolic $\psi_i$ functions are:
\begin{align}
\showlabel{hr2a}
\td{\psi_4}{\eta} & = \frac{\sinh \eta - \eta}{\cosh \eta - 1}, \\
\showlabel{hr2b}
\td{\psi_5}{\eta} & = \frac{\sinh \eta}{2} - \frac{3(\sinh \eta - \eta)}{2(\cosh \eta -1)}, \\
\showlabel{hr2c}
\td{\psi_6}{\eta} & = \frac{-1}{\cosh \eta - 1}.
\end{align}

%%%%%%%%%%%%%%%%%%%%%%%%%%%%%%%%%%%%%%%%%%%%%%%%%%%%%%%%%%%%%%%%%%%%%%%%%%%%%%%%%%%%%%%%%%%%%%%%%%%%
\subsection{Scaled Conformal Time}

%[[[Possibly move to numerical section???]]]

Below we will numerically calculate and plot results for a number of explicit models, and they will display a specific range of $t$ and $r$ values. The evolution calculations will be based on the conformal time $\eta$, so it will be necessary to set a suitable range of $\eta$ and to have sufficient data points in that range to make a good plot.  The relation between $t$ and $\eta$ depends on the scale time $\tilde{T}$, via \er{lt6}, \er{lt8} and \er{lt12}. In addition, near parabolic regions, where $f$ passes through zero, it can be seen that \er{lt6} and \er{lt8} imply $\eta\to0$.  To get a useful parameter in the parabolic case, the defintion $\overline{\eta} = \eta/\sqrt{|f|}$ is used before taking the limit $f \to 0$. These two problems are handled numerically by assuming a regular origin and defining the rescaled conformal time to be
\begin{align}
\showlabel{sct2}
\tilde{\eta} & = \frac{f^{1/2}}{M^{1/3}} \eta.
\end{align}

%% file: Section3.tex
%%%%%%%%%%%%%%%%%%%%%%%%%%%%%%%%%%%%%%%%%%%%%%%%%%%%%%%%%%%%%%%%%%%%%%%%%%%%%%%%%%%%%%%%%%%%%%%%%%%%
\section{Ricci Time Surfaces in LT Models}
\showlabel{Section3}

In the context of the LT model, we now investigate under what circumstances the surfaces of constant Ricci time, $S(\tau)$ are spacelike, null or timelike, for both hyperbolic and elliptic cases.  This involves comparing the slopes of the $S(\tau)$ surfaces with the slopes of the null rays.  Furthermore, we derive the necessary and sufficient conditions on the three arbitrary functions of the LT model, for the $S(\tau)$ to be timelike at early, late and intermediate times.  The results indicate the character of $\tau$ and provide a guide to construction numerical examples.

%%%%%%%%%%%%%%%%%%%%%%%%%%%%%%%%%%%%%%%%%%%%%%%%%%%%%%%%%%%%%%%%%%%%%%%%%%%%%%%%%%%%%%%%%%%%%%%%%%%%
\subsection{Ricci Time}
Ellis \cite{Ellis:2012ay} argued that the ``present" is a surface of constant time $S(\tau)$, and that the time $\tau$ for any $S(\tau)$ can be evaluated by integrating proper time along suitable worldlines, starting at the beginning of the universe up to $S(\tau)$. He proposed that the paths to use are the Ricci eigenlines, that is, the integral paths $x^a(\upsilon)$ of the timelike eigenvectors $\tilde{u}^a = dx^a/d\upsilon$ of the Ricci tensor:%
\footnote{
Eq \er{RicEV} follows from $R_{ab} = (T_{ab} - (1/2) T g_{ab})$ % \\
%$R_{ab} \tilde{u}^b = \lambda_2 \tilde{u}_a = (T_{ab} \tilde{u}^b - (1/2) T g_{ab} \tilde{u}^b)$ \\
%$R_{ab} \tilde{u}^b = \lambda_2 \tilde{u}_a = (\lambda_1 \tilde{u}_a - (1/2) T \tilde{u}_a)$ \\
~~$\to$~~ $\lambda_2 = \lambda_1 - (1/2) T$.
}
\begin{align}
\showlabel{RicEV}
T_{ab} \tilde{u}^b & = \lambda_1 \tilde{u}_a ~~~~~~\Leftrightarrow~~~~~~ R_{ab} \tilde{u}^b = \lambda_2 \tilde{u}_a, \\
x^a(\upsilon) & = \int_0^\upsilon \tilde{u}^a(\upsilon) \, d\upsilon \\
\tau & = \int_0^\upsilon \sqrt{|\tilde{u}^a \, g_{ab} \tilde{u}^b|}\; \, d\upsilon.
\end{align}
The resulting time $\tau$ will be called the Ricci time.  In the LT model, the Ricci eigenlines turn out to be the matter worldlines $u^a$, as is evident from \er{lt2}, and the proper time is just the local cosmic time since the bang,
\begin{align}
\showlabel{rt1}
\tau & =  t - a.
\end{align}
Consequently, the slope of $S(\tau)$ is simply the derivative of \er{rt1}, holding $\tau$ constant, i.e.
\begin{align}
\showlabel{rt2}
\left.\frac{dt}{dr}\right|_{\tau} & =  a'.
\end{align}
This already shows that the slope of the big bang surface is likely to be very significant.

%%%%%%%%%%%%%%%%%%%%%%%%%%%%%%%%%%%%%%%%%%%%%%%%%%%%%%%%%%%%%%%%%%%%%%%%%%%%%%%%%%%%%%%%%%%%%%%%%%%%
\subsection{Radial Light Rays \& the Spacelike/Timelike Conditions}
\showlabel{RLRSTC}
For radial null paths we have, $ds^{2} = d \theta^{2} = d \phi^{2}= 0$, and thus, \er{lt1} leads to
\begin{align}
\showlabel{stc1}
\left.\frac{dt}{dr}\right|_{n} & = \pm \frac{R'}{\sqrt{1+f}},
\end{align}
where the $+$ \& $-$ signs are used for outgoing and incoming paths respectively.  It is clear from section \ref{ParEvRr} that the appropriate expression for $R'$ depends on the value of $f$, so we must consider elliptic and hyperbolic cases seperately. The ratio of the null slope \er{stc1} to the slope of $S(\tau)$ \er{rt2} is%
\footnote{
We only need the ratio of absolute values, since the slopes of the incoming and outgoing light rays \er{stc1} have the same magnitude.  It is convenient to put the simpler expression in the denominator.
}
\begin{align}
\showlabel{stc1a}
\mathcal{R}  & =  \frac{|R'|}{|a' \sqrt{1+f}|}.
\end{align}
Locally, the surface $S(\tau)$ is timelike if it is steeper than the (radial) null path, and spacelike if it's less steep:
\begin{align}
& S(\tau)~\mbox{timelike:}~~~~~~~~~~~~ & 0 < \mathcal{R} & < 1 \nn \\
& S(\tau)~\mbox{null:}~~~~~~~~~~~~ & \mathcal{R} & = 1 \\
& S(\tau)~\mbox{spacelike:}~~~~~~~~~~~~ & 1 < \mathcal{R} & < \infty \nn
\end{align}
Generically, owing to \er{re2}, $R'/\sqrt{1 + f}\; \neq 0$ even at an extremum, only approaching it near the bang or crunch.  Therefore, a simultaneous bang, $a' = 0$, ensures $\cal R = \infty$ and $S(\tau)$ everywhere.  For all other considerations, the hyperbolic and elliptic cases need to be considered separately.

We wish to compare the location of any timelike $S(\tau)$ region with the horizon loci of the model.  The apparent horizon (AH) is the locus along a radial light ray where $R$ changes from increasing to decreasing.  Using equations \er{stc1} and \er{lt4} this gives
\begin{align}
\left.\td{R}{r}\right|_n = 0 = \dot{R} \left.\td{t}{r}\right|_n + R' = \dot{R}\left(\frac{\pm R'}{\sqrt{1+f}}\right) + R'
= \left(\frac{\mp\sqrt{2M/R+f}}{\pm\sqrt{1+f}}+1\right) R',
\end{align}
which leads to
\begin{align}
R_{AH} = 2M.
\end{align}
By equations \er{lt6} and \er{lt8}
\begin{align}
\showlabel{AHhyp}
\cosh\eta_{AH} & = 2f+1, & t_{AH} & = a+(\sinh\eta_{AH}-\eta_{AH})M/f^{3/2}, \\
\showlabel{AHell}
\cos(\eta_{AH}) & = 2f+1, & t_{AH} & = a+(\eta_{AH}-\sin\eta_{AH})M/(-f)^{3/2},
\end{align}
for hyperbolic and elliptic cases, respectively.

%%%%%%%%%%%%%%%%%%%%%%%%%%%%%%%%%%%%%%%%%%%%%%%%%%%%%%%%%%%%%%%%%%%%%%%%%%%%%%%%%%%%%%%%%%%%%%%%%%%%
\subsubsection{Hyperbolic Regions}

Substituting \er{lt10} into \er{stc1a}, we get
\begin{align}
\showlabel{stc4}
\mathcal{R} & = \frac{1}{|a'| \sqrt{1+f}} \bigg| \frac{M'}{f} \psi_4 + \frac{M f'}{f^2} \psi_5 + (-a') \sqrt{f} \psi_6 \bigg|.
\end{align}
The relative sizes of the $\psi_i$ functions varies greatly throughout the evolution, therefore we separate the model lifetime into three different regions, early, intermediate and late, and we investigate the ratio \er{stc4} in each of these regions.  
The early and late time behaviour of the various evolution functions $\psi_i$ were summarised in \S\ref{ParEvRr}.

\begin{itemize}

\item At early times ($\eta \rightarrow 0$), \er{stc4} reduces to
\begin{align}
\showlabel{hr1}
\mathcal{R} & \approx \frac{1}{|a'| \sqrt{1+f}}
\bigg| \frac{M'}{f} \frac{\eta^{2}}{6} + \frac{M f'}{f^5} \frac{\eta^4}{40} + (-a') \sqrt{f} \frac{2}{\eta} \bigg|
\approx \frac{2}{\eta} \sqrt{\frac{f}{1+f}},
\end{align}
assuming $a' \neq 0$, and it is clear that \er{hr1} goes to infinity.  This means that the surfaces $S(\tau)$ become spacelike at early times. Thus, for any $a'$, the $S(\tau)$ are spacelike everywhere at early enough times.

\item At intermediate times we have to deal with the full expression \er{stc4}, so only a rough indication can be obtained.  It makes sense to look for the minimum of $\mathcal{R}$ as $\eta$ changes, $\pdil{\cal R}{\eta} = 0$.  The derivatives of $\psi_{4}$, $\psi_{5}$ and $\psi_{6}$ are given in \er{hr2a}-\er{hr2c} and plotted in figure \ref{psi4,5,6min}.

\begin{figure}[!ht]
\centering 
\includegraphics[width=0.49\textwidth]{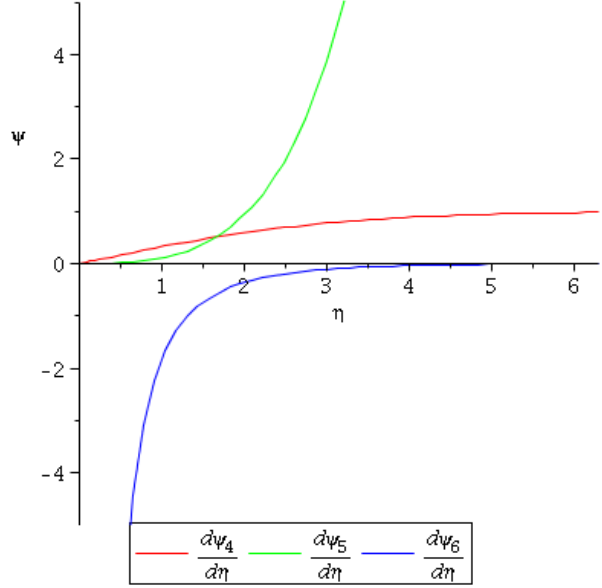}%
\caption{The shape of $\tdil{\psi_4}{\eta}$, $\tdil{\psi_5}{\eta}$ and $\tdil{\psi_6}{\eta}$ for the hyperbolic case.}
\showlabel{psi4,5,6min} 
\end{figure}

These $\eta$ functions are multiplied by the same functions of $r$ as in \er{stc4}, so the exact minimum $\eta$ will be different for each worldline.  But if for example we assume the $r$ functions are of similar magnitude, then $\eta_{min} \approx 1.7$. Now, \er{stc4} reduces to
\begin{align}
\showlabel{hr3}
\mathcal{R}  & = \frac{\big| A f M' + B M f' + C (-a') f^{5/2} \big|}{|a'| f^{2} \sqrt{1 + f}},
\end{align}
where $A = \psi_4(\eta_{min})$, $B = \psi_5(\eta_{min})$ and $C = \psi_6(\eta_{min})$ are constants.  At $\eta_{min} = 1.7$ they are $A = 0.46$, $B = 0.22$, $C = 1.45$.  For $S(\tau)$ to be timelike at $\eta = \eta_{min}$, $\cal R < 1$ in \er{hr3} leads to
\begin{align}
\showlabel{hr4a}
\sqrt{\frac{1+f}{f}} & > \bigg| \frac{\tilde{T}}{a'} \bigg( A\frac{M'}{M}+B\frac{f'}{f} \bigg) - C \bigg|,
\end{align}
where $\tilde{T}$ is the scale time of \er{lt12}.  It would appear to be quite easy to satisfy this constraint by making $|a'|$ large enough and $M'$ \& $f'$ small enough.

\item At late times ($\eta \rightarrow \infty$), \er{stc4} reduces to
\begin{align}
\showlabel{hr5}
\mathcal{R} & \approx
\frac{1}{|a'| \sqrt{1+f}} \bigg| \frac{M'}{f} \eta + \frac{M f'}{f^2} \frac{e^\eta}{4} + (-a') \sqrt{f}  \bigg|.
\end{align} 
If $a' \neq 0$, timelike $S(\tau)$ requires
\begin{align}
\showlabel{hr55}
M' = f' = 0,
\end{align}
in which case \er{hr5} reduces to $\mathcal{R} = \sqrt{f/(1+f)} \leq 1$, and this is always true, because $f > 0$.  Generally, $S(\tau)$ is timelike at late times only if $M' = f' = 0$ and $a' \neq 0$, otherwise it is spacelike.
\end{itemize}

%%%%%%%%%%%%%%%%%%%%%%%%%%%%%%%%%%%%%%%%%%%%%%%%%%%%%%%%%%%%%%%%%%%%%%%%%%%%%%%%%%%%%%%%%%%%%%%%%%%%
\subsubsection{Elliptic Regions}

Using \er{lt9}, equation \er{stc1a} evaluates to
\begin{align}
\showlabel{stc6}
\mathcal{R} & = \frac{1}{|a'| \sqrt{1+f}} \bigg| \frac{M'}{(-f)} \psi_1 + \frac{M f'}{f^2} \psi_2 + (-a') \sqrt{(-f)} \psi_3 \bigg|.
\end{align}

\begin{itemize}
\item At early times ($\eta \rightarrow 0$), \er{stc6} reduces to
\begin{align}
\showlabel{er1a}
\mathcal{R} & = \frac{1}{|a'| \sqrt{1+f}} \bigg| \frac{M'}{(-f)} \frac{\eta^2}{6} + \frac{M f'}{f^2} \frac{\eta^4}{40} + (-a') \sqrt{(-f)} \frac{2}{\eta} \bigg| \approx \frac{2}{\eta} \sqrt{\frac{(-f)}{1+f}} ,
\end{align}
assuming $a' \neq 0$. Clearly \er{er1a} diverges, which means $S(\tau)$ spacelike at early times for $a' \neq 0$, as well as $a' = 0$.

\item At an intermediate time, which we choose to mean maximum expansion, $\eta = \pi$, equation \er{stc6} reduces to (see table \ref{PsiLim})
\begin{align}
\showlabel{er4}
\mathcal{R} & =
\frac{|-2 M' f + 2 M f'|}{|a'| f^2\sqrt{1+f}} = \frac{1}{|a'| \sqrt{1+f}} \bigg| \frac{-2 M'}{f} + \frac{2 M f'}{f^2} \bigg|
= \frac{2 |\tilde{R}'|}{|a'|\sqrt{1+f}},
\end{align}
where $\tilde{R}'$ is the $r$ derivative of the scale length defined in \er{lt12}.  Applying the timelike condition $\mathcal{R} < 1$, \er{er4} gives
\begin{align}
\showlabel{er5}
|a'| & >  \frac{|\tilde{R}'|}{\sqrt{1+f}}.
\end{align}
As with the hyperbolic case, this is quite easy to satisfy on any given worldline.

\item At late times ($\eta = 2 \pi - \delta, \delta \rightarrow 0$), if $a' \neq 0$, equation \er{stc6} reduces to
\begin{align}
\showlabel{er7a}
\mathcal{R} & =
\frac{1}{|a'| \sqrt{1+f}} \Bigg| \frac{M'}{(-f)} \frac{4 \pi}{\delta} + \frac{M f'}{f^{2}} \frac{6 \pi}{\delta} + (-a') \sqrt{(-f)} \bigg(\frac{-2}{\delta} \bigg) \Bigg|, \nonumber \\
& = \frac{2 \big|2\pi\tilde{T'} + a'\big|}{|a'| \delta} \sqrt{\frac{(-f)}{1+f}},
\end{align}
$\tilde{T}'$ being the $r$ derivative of the scale tine in \er{lt12}.  It is clear that $\mathcal{R}$ diverges unless $2\pi\tilde{T'} + a'= 0$, which makes $\mathcal{R} = 0$ and $S(\tau)$ a timelike surface.  We find that $S(\tau)$ becomes timelike near the big crunch only if the crunch time is constant and the bang time is not.
\end{itemize}

Finally, we can summarize these results in table \ref{Ricci}.
%\begin{landscape}
\begin{table}[!h]
\begin{center}
\begin{tabular}{|l|l|l|l|}
\hline \hline
      & Early times & Intermediate times
           & Late times \\ \hline \hline
Hyperbolic
      & \pb{16mm}{Always spacelike}
           & \pb{60mm}{$S(\tau)$ timelike only if
             $$\sqrt{\frac{1+f}{f}} > \Bigg| \frac{\tilde{T}}{a'} \bigg( A\frac{M'}{M}+B\frac{f'}{f} \bigg) - C\Bigg|$$
             and $a' \neq 0$, otherwise spacelike}
                & \pb{40mm}{$S(\tau)$ timelike if
                  $$M' = f' = 0$$
                  and $a' \neq 0$, otherwise spacelike} \\ \hline \hline
Elliptic
      & \pb{16mm}{Always spacelike}
           & \pb{60mm}{$S(\tau)$ timelike only if
             $$|a'| >  \frac{|\tilde{R}'|}{\sqrt{1+f}}$$
             and $a' \neq 0$, otherwise spacelike}
                & \pb{40mm}{$S(\tau)$ timelike if $$ 2\pi\tilde{T}' + a' = 0$$
                  and $a' \neq 0$, otherwise spacelike} \\ \hline \hline
\end{tabular}
\pb{160mm}{\caption{Summary of the conditions for $S(\tau)$ to be timelike or spacelike at early, middle and late times.  It is understood that $S(\tau)$ is null on the boundary between timelike and spacelike regions.}
\showlabel{Ricci}}
\end{center}
\end{table}
%\end{landscape}

%% file: Section4.tex
%%%%%%%%%%%%%%%%%%%%%%%%%%%%%%%%%%%%%%%%%%%%%%%%%%%%%%%%%%%%%%%%%%%%%%%%%%%%%%%%%%%%%%%%%%%%%%%%%%%%
\section{Explicit Models}
\showlabel{Section4}

We considered several different time regimes of \LT\ models in section \ref{Section3}, and derived the necessary and sufficient conditions for the constant Ricci time surfaces, $S(\tau)$, to be timelike or spacelike.  The main purpose of this section is to investigate numerically the detailed behaviour of these surfaces in a range of different explicit LT models, including elliptic and hyperbolic cases, and thereby visualize the character of $S(\tau)$ in $(t,r)$ plots.  We also check if the regions of timelike $S(\tau)$ occur inside or outside the apparent horizons.  The models considered below are only intended to be realistic over the finite range of $r$ that is plotted.

%%%%%%%%%%%%%%%%%%%%%%%%%%%%%%%%%%%%%%%%%%%%%%%%%%%%%%%%%%%%%%%%%%%%%%%%%%%%%%%%%%%%%%%%%%%%%%%%%%%%
\subsection{Numerical Calculations}
We have developed a numerical code that takes the definitions of the LT arbitrary functions and calculates $\tdil{t}{r}|_n$, $\tdil{t}{r}|_\tau$ and $\mathcal{R}$, as detailed in section \ref{Section3}. The resulting data is plotted against $r$ and $t$ as a pair of surfaces of different colours, whose heights above the $(t,r)$ plane are the slopes $\tdil{t}{r}|_n$ and $\tdil{t}{r}|_\tau$.  In order to check that our model choices were reasonable and free of irregularities, the code also produces plots of $M(r)$, $f(r)$, $a(r)$ and $\tilde{T}(r)$, as well as the functions that indicate a shell crossing if they have the wrong sign, $M'(r)$, $f'(r)$, $a'(r)$ and $2\pi\tilde{T}'(r)+a'(r)$.  Out of these last 4 functions, the first 3 are relevant to hyperbolic regions, and functions 1, 3 and 4 are relevant to elliptic models (see table \ref{S}).

The actual code must allow for a few practical numerical issues.
\begin{itemize}
\item The slope $\tdil{t}{r}|_n$ diverges near the bang and the crunch, and this must be chopped by not plotting certain $\eta$ values.
\item In elliptic regions, if the big crunch occurs within the time range plotted, the two surfaces must be terminated just before the crunch.
\item In calculating the locus of the AH, it is easiest to use \er{AHhyp} \& \er{AHell}, since we are already using $\eta$ in the evolution calcuations.  In fact there are two AHs in elliptic regions.  Although each AH is a locus in the $(t,r)$ plane, for clarity, the AH curves are plotted on the null slope surfaces, i.e. at ``height" $\tdil{t}{r}|_n$ on the plot, otherwise they would be hidden.
\item As noted above, the origin and spatial extrema in $R$ generate undefined zero-over-zero values unless their limits are evaluated properly.  Therefore, the various arbitrary funtion combinations appearing in equations \er{stc4} and \er{stc6}, such as $M f'/f^2/\sqrt{1+f}/a'$ and $\sqrt{f/(1+f)}$, must be re-evaluated analytically for each choice of $M$, $f$ and $a$, and written into the code explicitly.
\end{itemize}

The general algorithm is as follows.
\begin{enumerate}
 \item Define the ranges of $r$ and $\eta$ (typically $r$ changes from $0$ to $0.9$, while $\eta$ extends from $0$ to $7.2$).
 \item Define the arbitrary functions $M$, $f$, $a$ and their derivatives $M'$, $f'$, $a'$, as well as other ratios like $M/f$ and $M/f^{3/2}$. 
 \item Rescale $\eta$ to be $\tilde{\eta}$ given by \er{sct2} in order to avoid vastly different timescales in the same plot. and handle near-parabolic regions without trouble.
 \item In terms of $\tilde{\eta}$, calculate the $\psi_i$ for elliptic regions from \er{lt9a}, \er{lt9b} and \er{lt9c}, and for hyperbolic regions from \er{lt10a}, \er{lt10b} and \er{lt10c}.  Redefine these functions as series approximations for use when $\eta$ approaches zero or $2 \pi$.
 \item Depending on the value of $\tilde{\eta}$, calculate $R$, $t$, $dt/dr|_n$, $dt/dr|_\tau$ and $\mathcal{R}$ for the whole range, using the appropriate elliptic or hyperbolic expressions.
 \item Plot the two slopes as intersecting 2-d surfaces in a 3-d plot.\footnote{Several other plots, such as $\mathcal{R}$ against $t$ and $R$, are plotted but not used here.}
 \item Evaluate the AH locus, and draw it as a black line on the $dt/dr|_n$ surface.
 \item Plot the LT arbitrary functions and the no shell crossing condition functions.
\end{enumerate}

%%%%%%%%%%%%%%%%%%%%%%%%%%%%%%%%%%%%%%%%%%%%%%%%%%%%%%%%%%%%%%%%%%%%%%%%%%%%%%%%%%%%%%%%%%%%%%%%%%%%
\subsection{Hyperbolic Models}
%%%%%%%%%%%%%%%%%%%%%%%%%%%%%%%%%%%%%%%%%%%%%%%%%%%%%%%%%%%%%%%%%%%%%%%%%%%%%%%%%%%%%%%%%%%%%%%%%%%%
\subsubsection{Model 1}
The arbitrary functions
\begin{align}
\showlabel{hm11}
M & = M_{0}(r^{3} + M_{1}r^{4} + M_{2}r^{5}),\\
\showlabel{hm12}
f & = - k (r^{2} + f_{1}r^{3} + f_{2}r^{4}),\\
\showlabel{hm13}
a & = a_{0} + a_{1}r + a_{2}r^{2} + a_{3}r^{3},
\end{align}
describe an inhomogeneous model with non-zero density everywhere, so it could be part of a ``cosmology".  The first set of coefficients is $M_{0} = 1$, $M_{1} = -3$, $M_{2} = 2.4$, $k = - 1$, $f_{1} = -2.67$, $f_{2} = 2$, $a_{0} = 5$, $a_{1} = -200$, $a_{2} = 0.5$ and $a_{3} =  0.9$.  These functions and coefficient values were specifically selected so that both $M$ and $f$ have zero gradient at the same $r$ value, but never have a negative gradient, as shown in figure \ref{hmodel11}.  Therefore we have a worldine ($r$ value) which satisfies the condition \er{hr55}, $M' = f' = 0$, which should make $S(\tau)$ timelike at late times, and the neighbouring worldlines are very close to that condition.
Figure \ref{hmodel11} plots the arbitrary functions $M$, $f$ and $a$.  Scale time $\tilde{T}$ is not necessarily increasing with $r$, but has a zero gradient at the same point as functions $M$ and $f$.  Figure \ref{hmodel12} shows the conditions for no shell crossings are satisfied, because $M'\geq0$, $f'\geq0$ and $a'\leq0$ clearly hold.  The fourth condition $2 \pi \tilde{T}^{'} + a^{'} \geq 0$ is not relevant for hyperbolic models.

\begin{figure}[!hb]
\centering
\subfloat[The behavior of $M$, $f$, $a$ and $\tilde{T}$\showlabel{hmodel11}]{\includegraphics[width=0.49\textwidth]{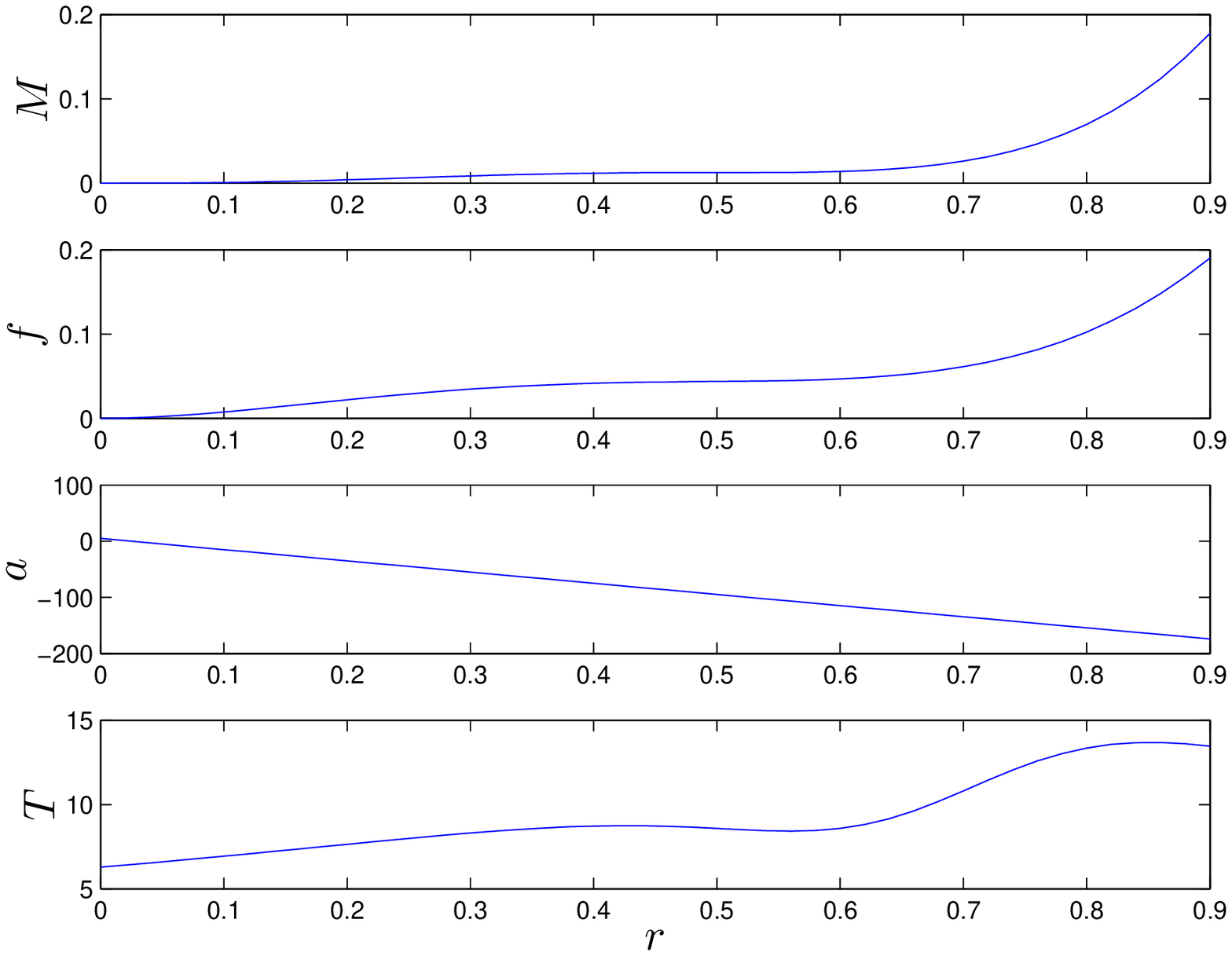}}
\subfloat[Conditions for no shell crossings\showlabel{hmodel12}]{\includegraphics[width=0.5\textwidth]{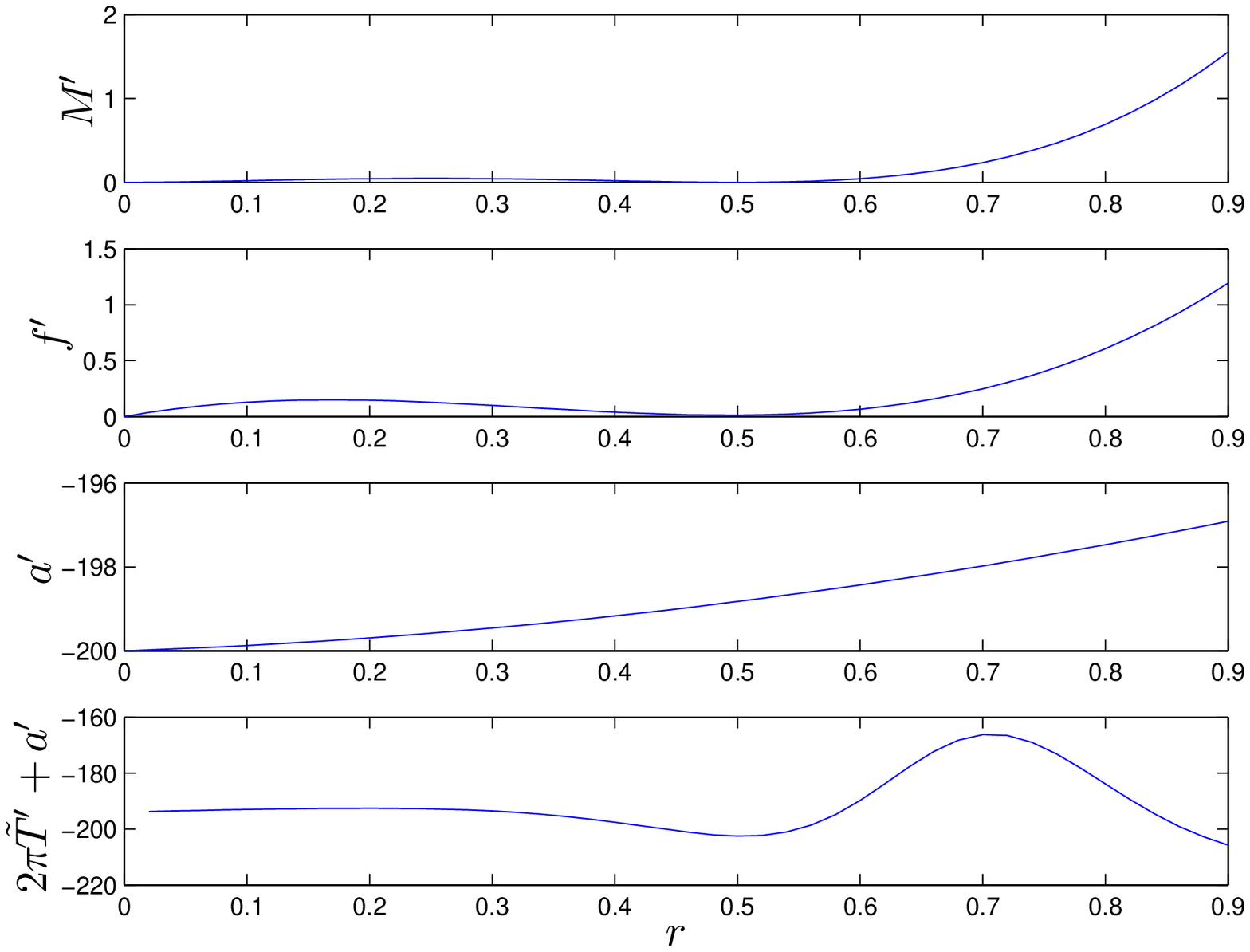}}
\caption{Arbitrary Functions for Hyperbolic model (1) using the first set of coefficients.  The model is free of shell crossings if the $M'$ and $f'$ graphs are never negative, and the $a'$ graph is never positive.  See table \ref{S}. \showlabel{hmodel1}}
\end{figure}

\begin{figure}[!hb]
\centering 
\includegraphics[width=0.68\textwidth]{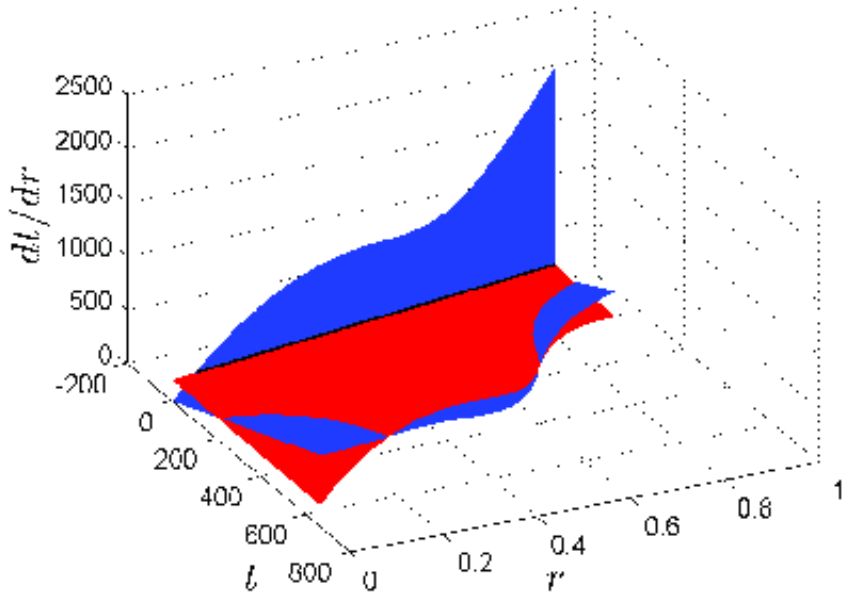}
\caption{Hyperbolic model (1) using the first set of coefficients: The slopes of the $S(\tau)$ and radial null surfaces, $dt/dr|_\tau$ and $dt/dr|n$, shown in red and blue respectively, are plotted against $t$ and $r$.  Where the red surface is above the blue one, $S(\tau)$ is timelike. The black line represents the apparent horizon, and the big bang is where the blue surface rises at the back.\showlabel{hmodel13}}
\end{figure}

Figure \ref{hmodel13} illustrates the relation between the \{constant $\tau$\} surfaces $S(\tau)$ (red) and the radial null surfaces (blue) for each worldline and a sufficiently large range of the evolution.  The black line shows the apparent horizon, which allows us to see whether the timelike $S(\tau)$ regions occur inside (at smaller $R$) or outside (at larger $R$) the AH. Our hyperbolic model is always expanding, so smaller $R$ occurs to the past of the black line.  For the worldlines $(0.3<r<0.7)$, the $S(\tau)$ are spacelike at early times, they become timelike at intermediate times and remain so for the rest of their plotted evolution.  This confirms that $S(\tau \to \infty)$ is timelike on $r = 0.5$ for which $M' = f' = 0$.  For the remaining worldlines, $(0<r<0.3)$ and $(0.3<r<0.7)$, the $S(\tau)$ are spacelike at early and late times, but change to timelike for at least part of the evolution. We noticed that the size (duration) of the intermediate timelike region, can be increased by increasing $(-a')$. Conversely, decreasing $(-a')$ makes it shrink towards the bang.  This model confirms that $S(\tau)$ is spacelike everywhere for early times, but at late times $S(\tau)$ can be timelike only if $M'$ and $f'$ are zero while $a'$ is not.  It also illustrates that timelike $(\tau)$ at intermediate times is not hard to achieve.  The black line indicates that the timelike region is located outside the AH, i.e. at larger $R$. 

\begin{figure}[!hb]
\centering
\subfloat[The behavior of $M$, $f$, $a$ and $\tilde{T}$\showlabel{hmodel31}]{\includegraphics[width=0.49\textwidth]{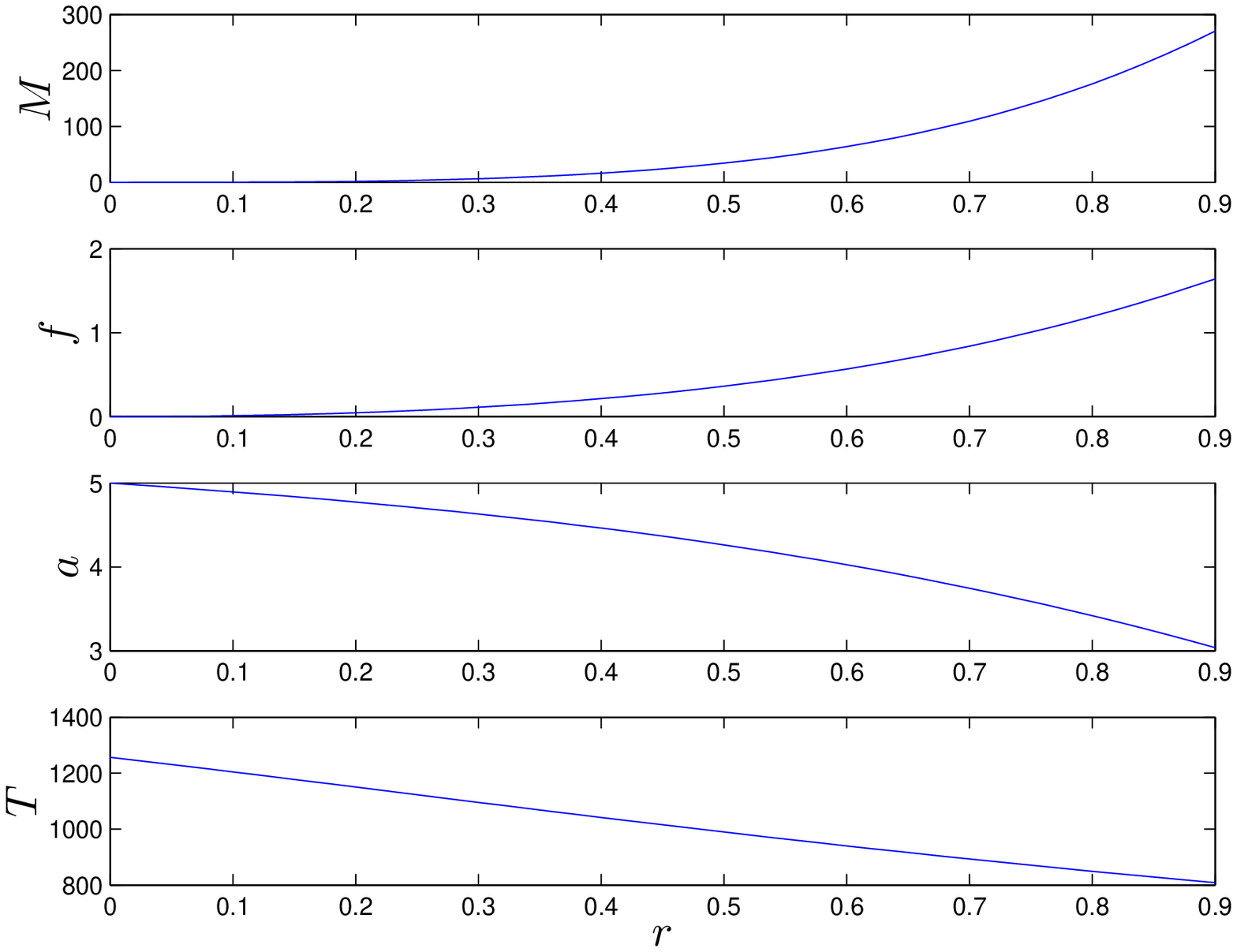}}
\subfloat[Conditions for no shell crossings\showlabel{hmodel32}]{\includegraphics[width=0.5\textwidth]{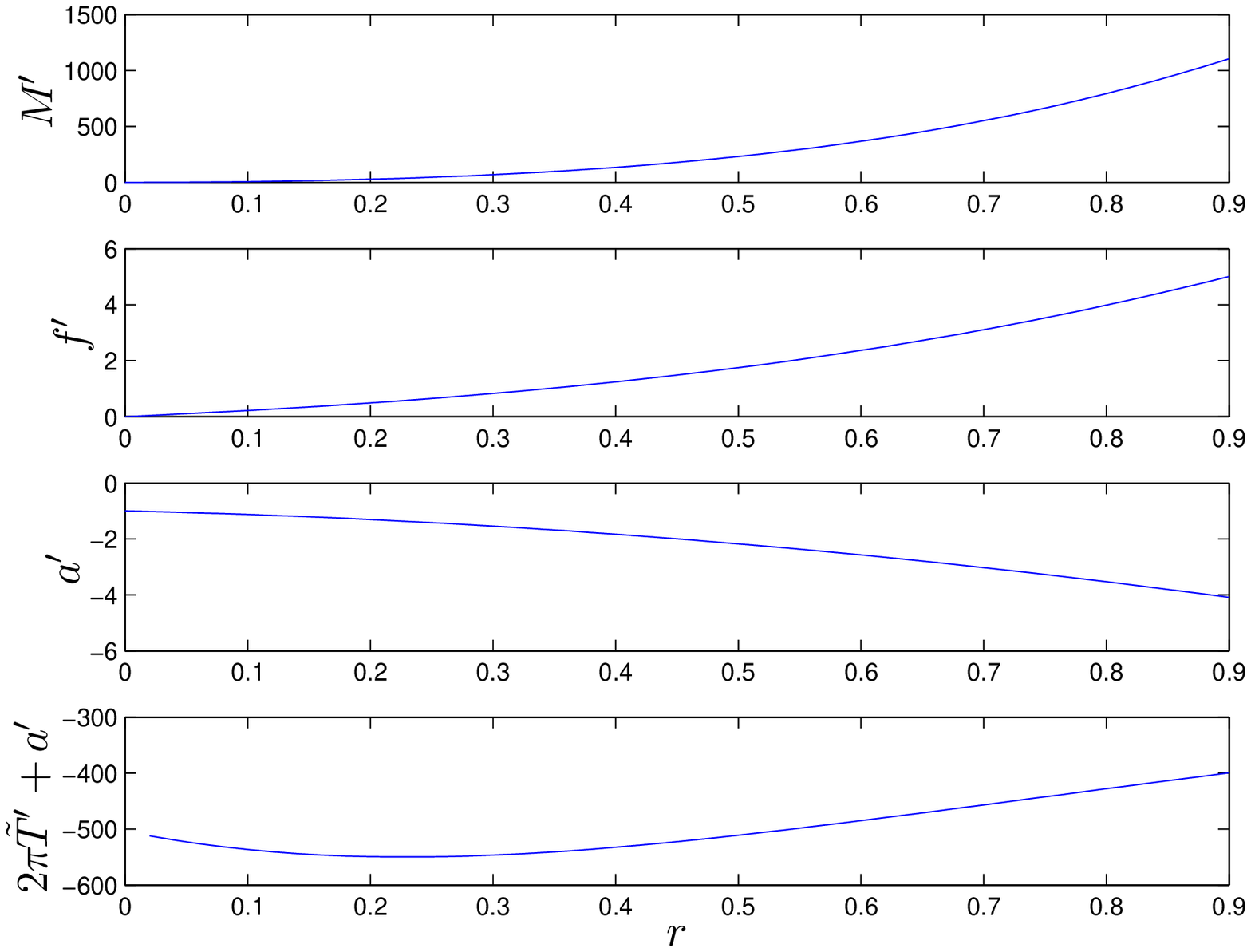}}
\caption{Arbitrary Functions for Hyperbolic model (1) using the second set of coefficients.\showlabel{hmodel3}}
\end{figure}

Now, by changing the coefficients of the arbitrary functions in equations \er{hm11}-\er{hm13}, we can produce a model for which $S(\tau)$ is spacelike everywhere. Taking $M_{0} = 200$, $M_{1} = 0.5$, $M_{2} = 0.5$, $k = - 1$, $f_{1} = 0.6$, $f_{2} = 0.6$, $a_{0} = 5$, $a_{1} = -1$, $a_{2} = -0.5$ and $a_{3} = -0.9$, we get the LT functions shown in figure \ref{hmodel31}, and figure \ref{hmodel32} shows there are no shell crossings.  The blue $\tdil{t}{r}|_n$ surface in figure \ref{hmodel33} is everywhere above the red $\tdil{t}{r}|\tau$ surface, showing that $S(\tau)$ is always spacelike for each world line.  This is because $M'>0$ and $f'>0$ and $a'$ is not so large. The bold black line visualizes the apparent horizon.

\begin{figure}[!hb]
\centering 
\includegraphics[width=0.7\textwidth]{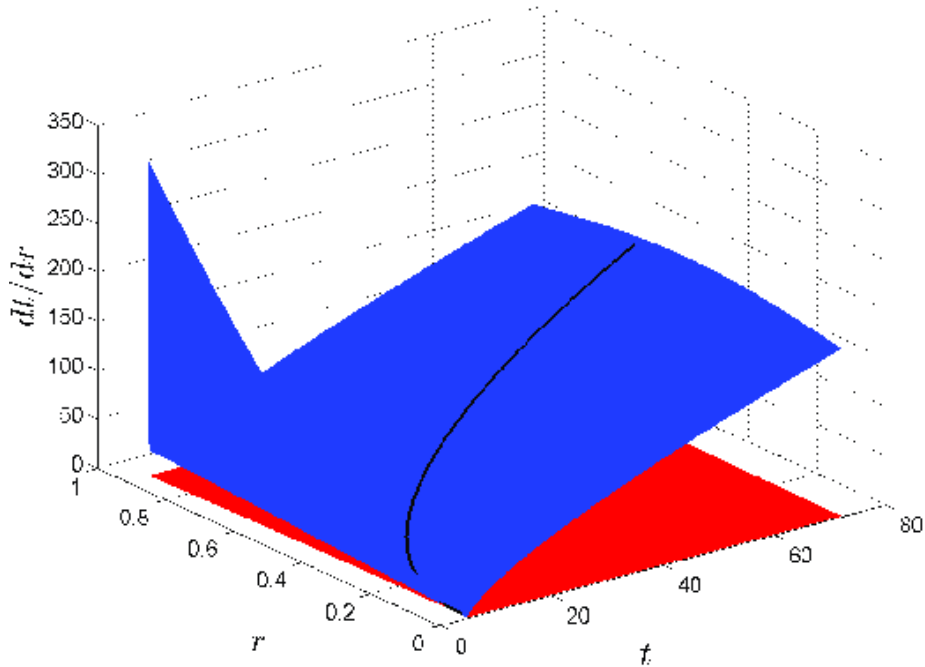}
\caption{Hyperbolic model (1) using the second set of coefficients: The slopes of the $S(\tau)$ and radial null surfaces are plotted, showing $S(\tau)$ is spacelike everywhere.\showlabel{hmodel33}}
\end{figure}

By using a third set of arbitrary functions and varying just one of them, we can investigate the nature of the intermediate time. Choosing $M_{0} = 1$, $M_{1} = -2$, $M_{2} = 1.4$, $k = - 1$, $f_{1} = -1.65$, $f_{2} = 1$, $a_{0} = 0$, $a_{1} = -200$, $a_{2} = 0.1$ and $a_{3} =  0.1$, the behavior of the functions and the conditions for no shell crossings are represented in Figures \ref{hmodel41} and \ref{hmodel42}, respectively. The coefficient $a_1$ will be varied.

\begin{figure}[!hb]
\centering
\subfloat[The behavior of $M$, $f$, $a$ and $\tilde{T}$\showlabel{hmodel41}]{\includegraphics[width=0.47\textwidth]{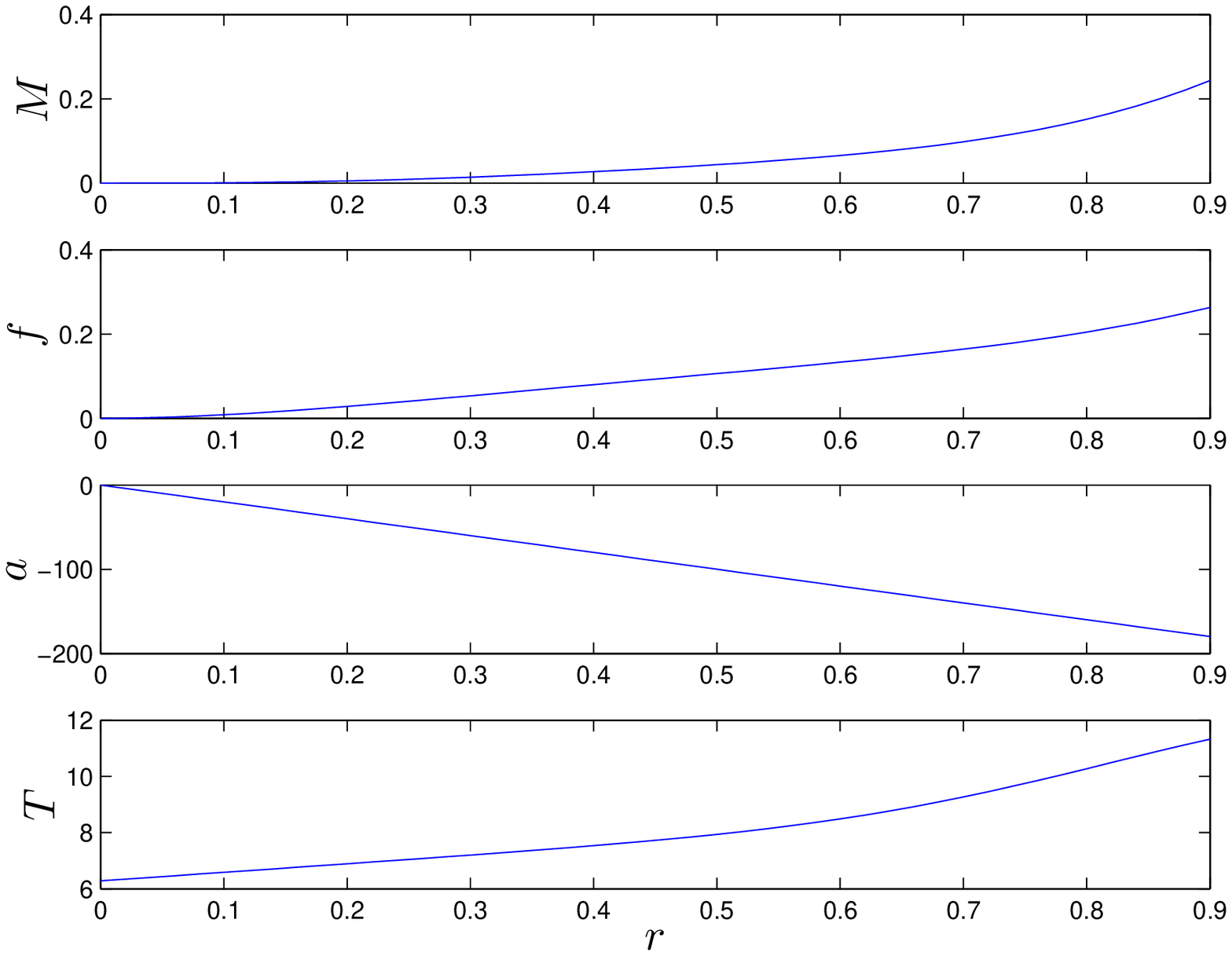}}
\subfloat[Conditions for no shell crossings\showlabel{hmodel42}]{\includegraphics[width=0.49\textwidth]{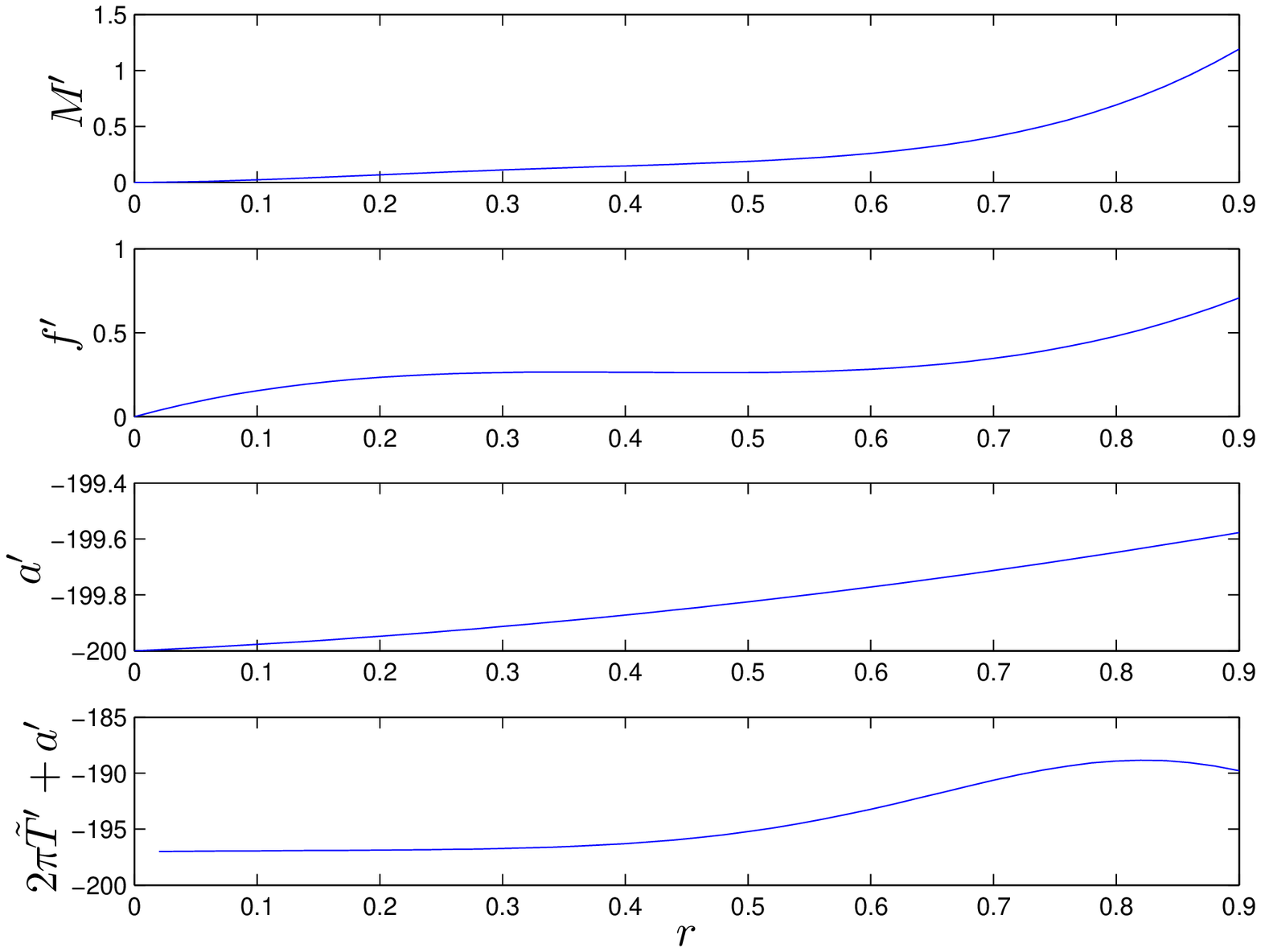}}
\caption{Arbitrary Functions for Hyperbolic model (1) using the third set of coefficients and $a_1=-200$.\showlabel{hmodel4}}
\end{figure}

\begin{figure}[!hb]
\centering 
\includegraphics[width=0.7\textwidth]{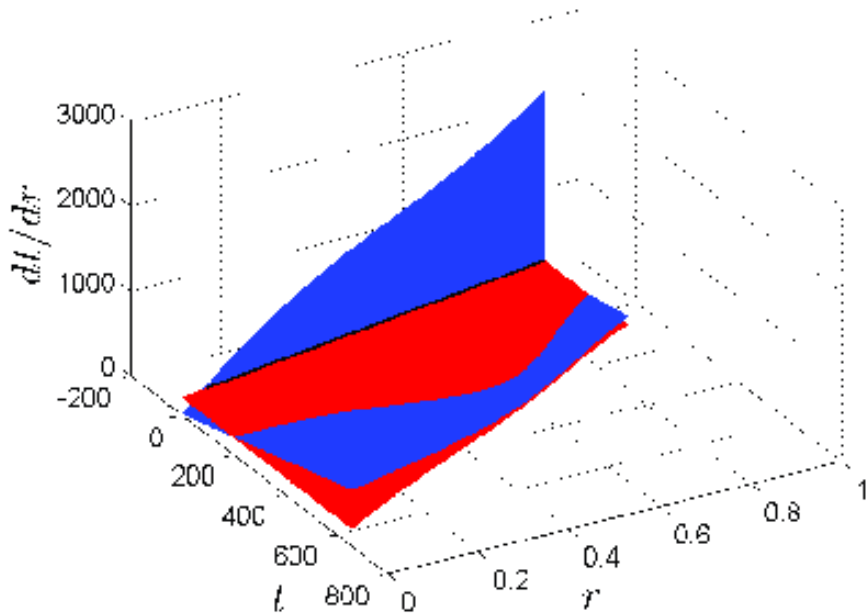}
\caption{Hyperbolic model (1) using the third set of coefficients and $a_1=-200$: The slopes of the $S(\tau)$ and null surfaces are shown, and here $S(\tau)$ is timelike only at intermediate times.\showlabel{hmodel43}}
\end{figure}

With $a_1=-200$, which creates a moderately large value of $(-a')$, the $S(\tau)$ are spacelike for all $r$, at both early and late times, but they still become timelike at intermediate times, figure \ref{hmodel43}. If $(-a')$ is made very large, $a_{1}=-2000$, the $S(\tau)$ are of course spacelike at the earliest times, but become timelike just after the AH, and remain timelike up to very late times on all worldlines, figure \ref{h41}.  However, since $M'>0$ and $f'>0$, the $S(\tau)$ must eventually become spacelike again.  If $(-a')$ is made small, $a_{1}=-0.2$, the $S(\tau)$ now become spacelike everywhere at all times, figure \ref{h43}.  In other words, by decreasing $(-a')$, the extent of the timelike region can be decreased and made to vanish for small enough $(-a')$. Moreover, the timelike region shrinks towards the origin as $(-a')$ is decreased.

\begin{figure}[!hb]
\centering
\subfloat[$a_{1}$ = -2000\showlabel{h41}]{\includegraphics[width=0.33\textwidth]{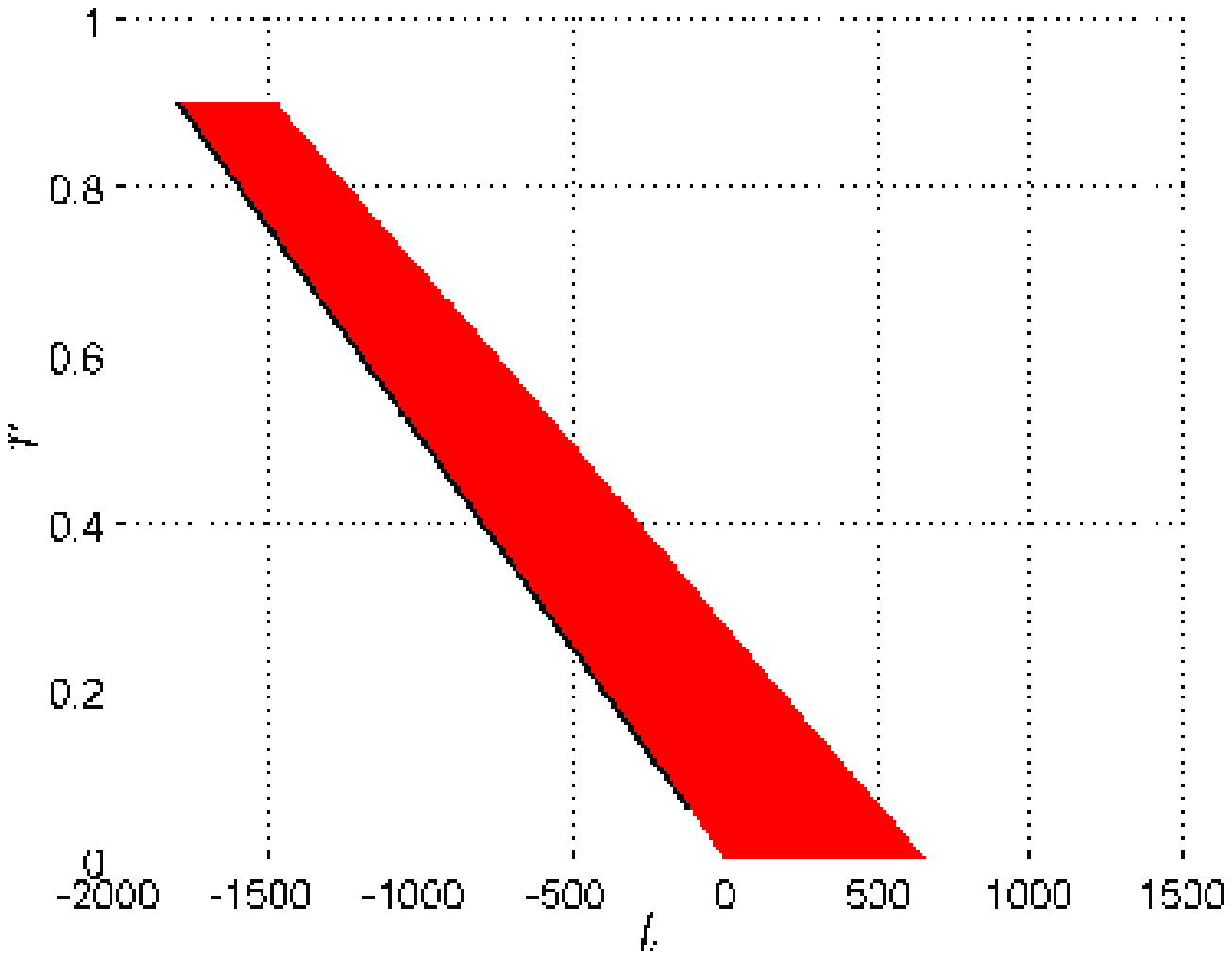}}
\subfloat[$a_{1}$ = -200\showlabel{h42}]{\includegraphics[width=0.33\textwidth]{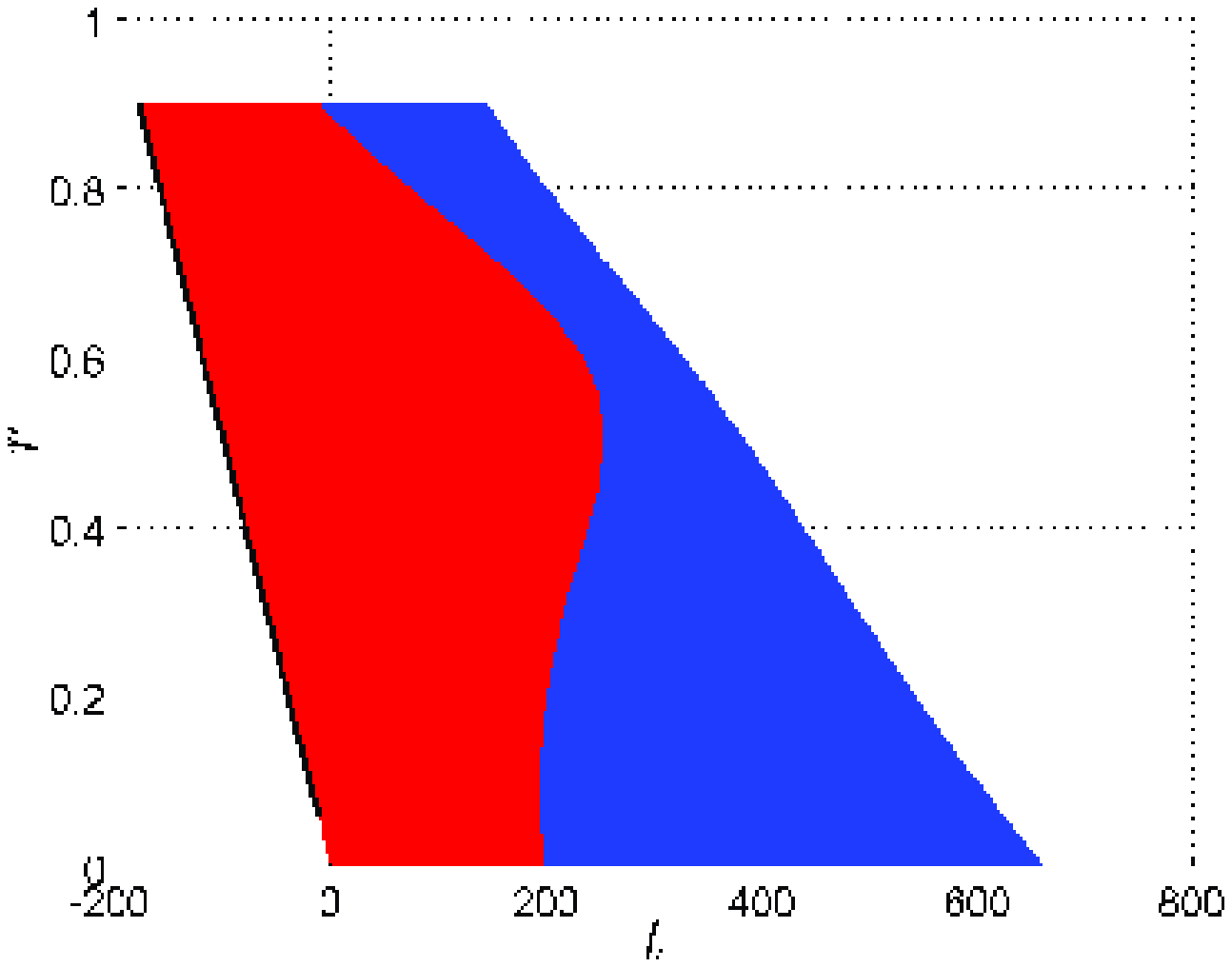}}
\subfloat[$a_{1}$ = -0.2\showlabel{h43}]{\includegraphics[width=0.33\textwidth]{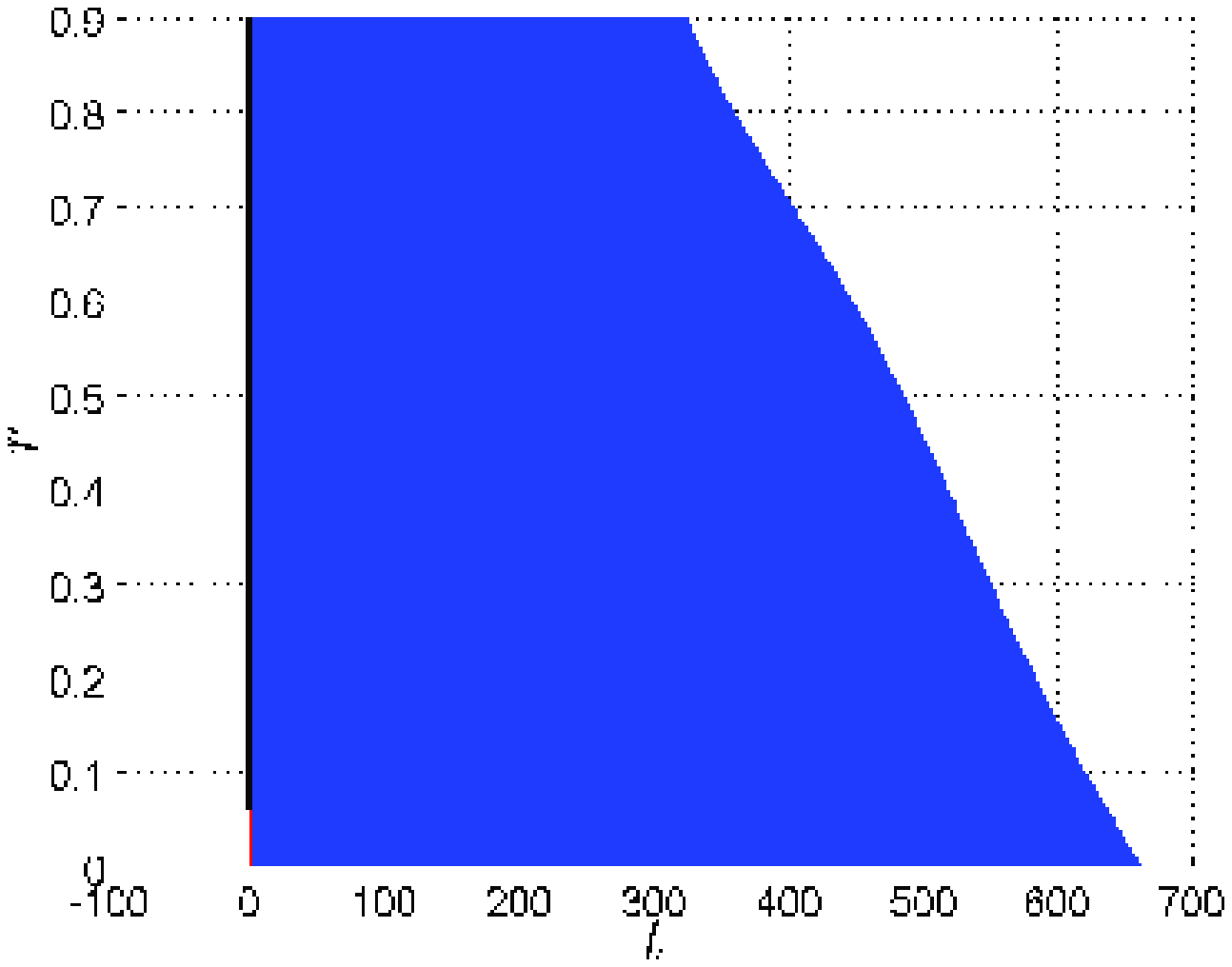}}
\caption{Hyperbolic model (1) using the third set of coefficients and 3 different $a_1$ values: This is a view looking straight ``down", so the ``heights" $\tdil{t}{r}$ are not seen.  The sizes of the red regions show the sizes of the regions of timelike $S(\tau)$.  The black line is the apparent horizon.\showlabel{h4}}
\end{figure}

%%%%%%%%%%%%%%%%%%%%%%%%%%%%%%%%%%%%%%%%%%%%%%%%%%%%%%%%%%%%%%%%%%%%%%%%%%%%%%%%%%%%%%%%%%%%%%%%%%%%
\subsubsection{Model 2}
The arbitrary functions are 
\begin{align}
\showlabel{hm21}
M & = M_{0} ~{\rm exp}(1 + M_{1}r),\\
\showlabel{hm22}
f & = - k ~{\rm exp}(1 + f_{1}r),\\
\showlabel{hm23}
a & = a_{0} ~{\rm exp}(1 + a_{1}r),
\end{align}
where $M_{0} = 2$, $M_{1} = 0.01$, $k = - 1$, $f_{1} = 0.1$, $a_{0} = -0.01$ and $a_{1} = 5$, and they're plotted in figure \ref{hmodel21}.  This is a hyperbolic model with an unusual topology, because the ``origin" is asymptotically far away at $r=-\infty$.  However, figure \ref{hmodel22} shows it is free of shell crossings.

\begin{figure}[!hb]
\centering
\subfloat[The behavior of $M$, $f$, $a$ and $\tilde{T}$\showlabel{hmodel21}]{\includegraphics[width=0.49\textwidth]{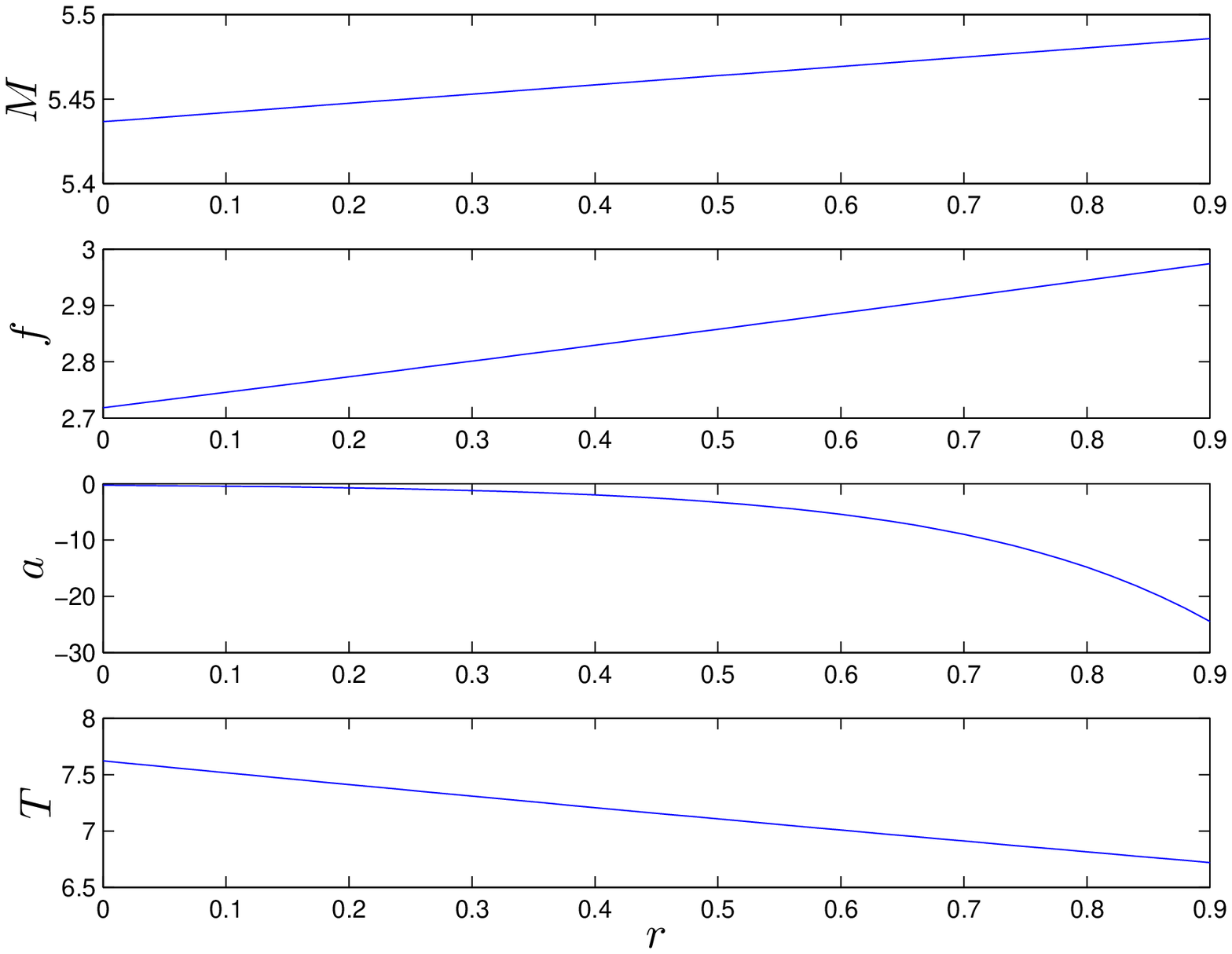}}
\subfloat[Conditions for no shell crossings\showlabel{hmodel22}]{\includegraphics[width=0.5\textwidth]{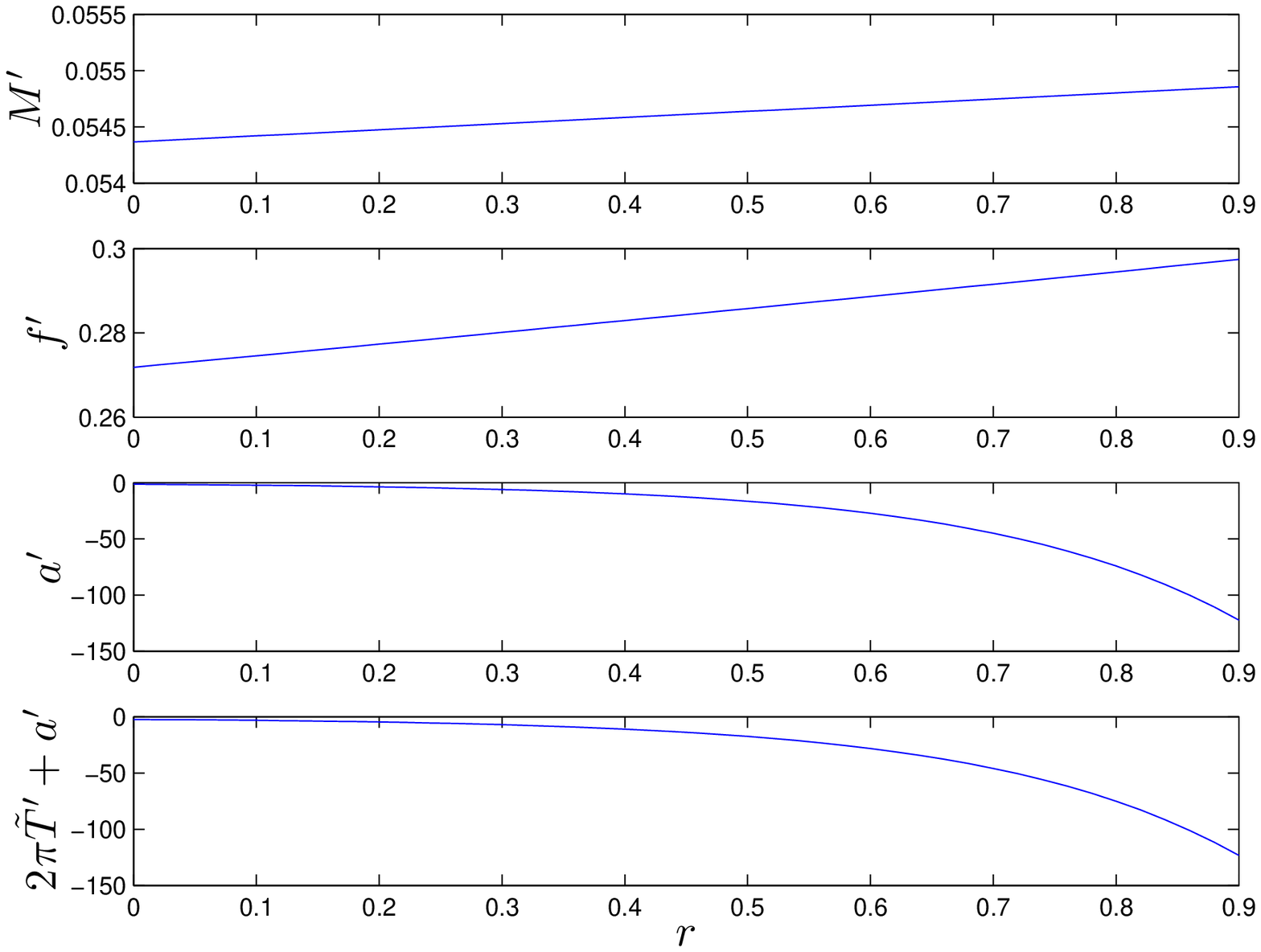}}
\caption{Arbitrary Functions for Hyperbolic model (2).\showlabel{hmodel2}}
\end{figure}

\begin{figure}[!hb]
\centering 
\includegraphics[width=0.7\textwidth]{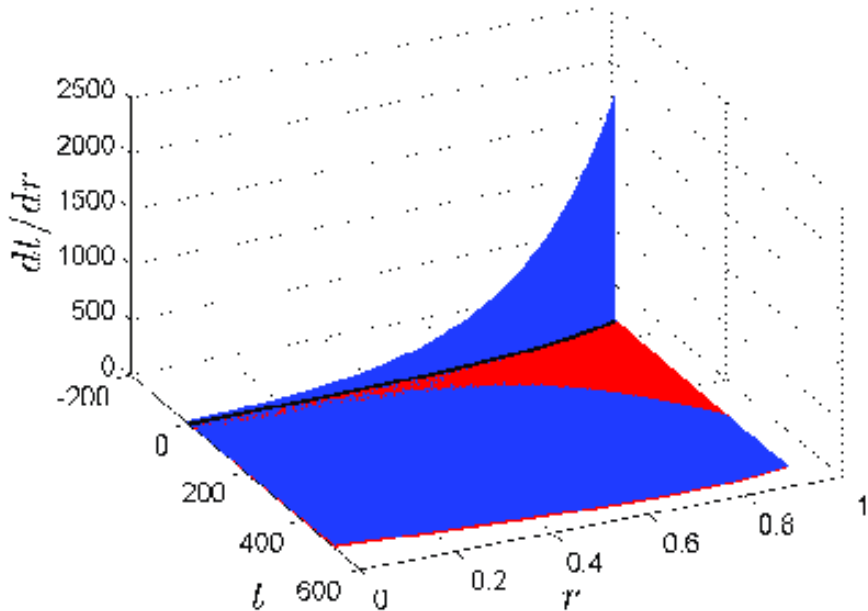}
\caption{Hyperbolic model (2): The slopes of the $S(\tau)$ and null surfaces are indicated by the heights of the red and blue surfaces in this plot, and $S(\tau)$ is timelike where the red surface is higher.\showlabel{hmodel23} }
\end{figure}

At early times, the $S(\tau)$ are spacelike everywhere, but become timelike at intermediate times for a specific region. As before, the size of the timelike region depends strongly on $(-a')$, and it occurs outside the AH. At late times, the $S(\tau)$ eventually become spacelike again.

%%%%%%%%%%%%%%%%%%%%%%%%%%%%%%%%%%%%%%%%%%%%%%%%%%%%%%%%%%%%%%%%%%%%%%%%%%%%%%%%%%%%%%%%%%%%%%%%%%%%
\subsection{Elliptic Models}
%%%%%%%%%%%%%%%%%%%%%%%%%%%%%%%%%%%%%%%%%%%%%%%%%%%%%%%%%%%%%%%%%%%%%%%%%%%%%%%%%%%%%%%%%%%%%%%%%%%%
\subsubsection{Model 1}
The following arbitrary functions
\begin{align}
\showlabel{em11}
M & = M_{0}(r^{3} + M_{1}r^{4} + M_{2}r^{5}),\\
\showlabel{em12}
f & = - k (r^{2} + f_{1}r^{3} + f_{2}r^{4}),\\
\showlabel{em13}
a & = a_{0} + a_{1}r + a_{2}r^{2} + a_{3}r^{3},
\end{align}
are first considered with $M_{0} = 1$, $M_{1} = 0$, $M_{2} = 2.4$, $k = 1$, $f_{1} = -0.67$, $f_{2} = 0$, $a_{0} = 5$, $a_{1} = -5$, $a_{2} = 0.5$ and $a_{3} =  0.9$, as plotted figure \ref{emodel11}.  This function set represents an inhomogeneous elliptic model, which could be part of a cosmology.  Although these functions would not work globally, they are entirely valid for the range of $r$ calculated.  Figure \ref{emodel12} shows no shell crossings occur; for elliptic models the $f'$ plot is not relevant.

\begin{figure}[!hb]
\centering
\subfloat[The behavior of $M$, $f$, $a$ and $\tilde{T}$\showlabel{emodel11}]{\includegraphics[width=0.49\textwidth]{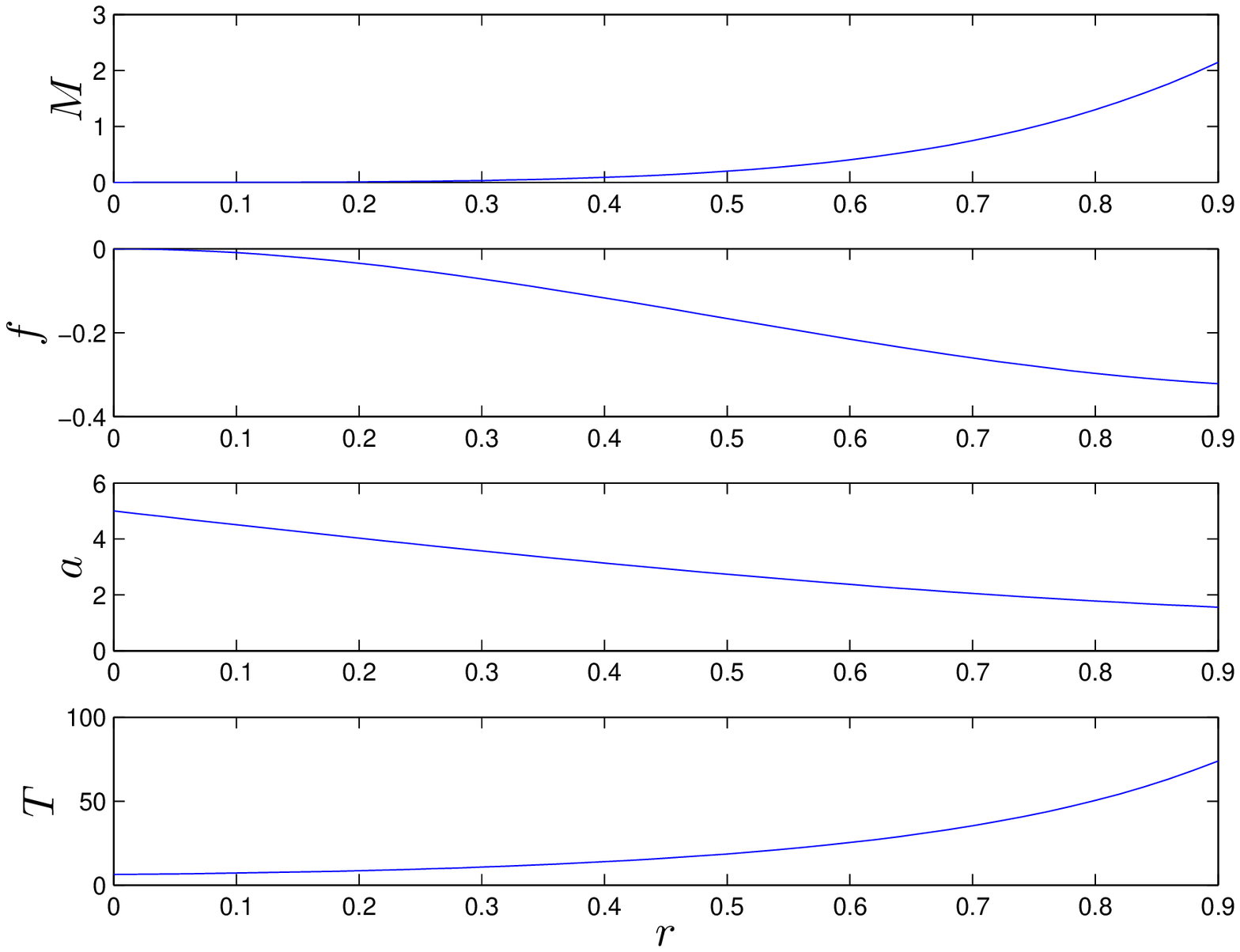}}
\subfloat[Conditions for no shell crossings\showlabel{emodel12}]{\includegraphics[width=0.5\textwidth]{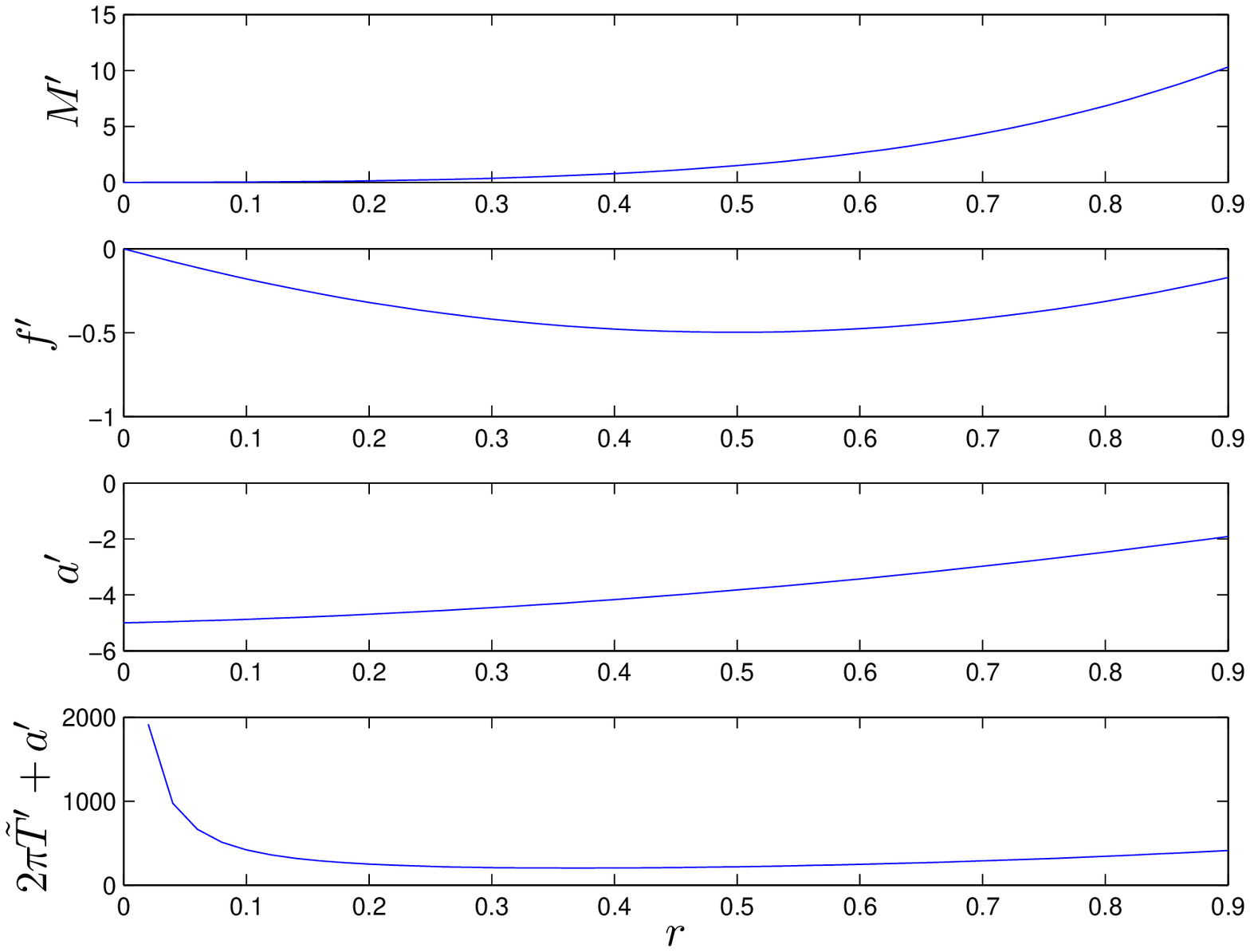}}
\caption{Arbitrary Functions for Elliptic model (1) with the first coefficient set.  The model is free of shell crossings if the $M'$ and $2 \pi \tilde{T}' + a'$ graphs are never negative, and the $a'$ graph is never positive.  See table \ref{S}. \showlabel{emodel1}}
\end{figure}

Figure \ref{emodel13} illustrates the relationship between the $S(\tau)$ and the null surfaces, whose slopes are represented by the heights of the red and blue surfaces respectively, in the 3-d plot.  The bold black lines show the loci of the apparent horizons.  At early and late times, the $S(\tau)$ surfaces are spacelike everywhere, but the $S(\tau)$ become timelike at intermediate times, for a range of worldlines $(0<r<0.3)$.  The extent of the timelike region once again depends on the gradient of the bang surface, and as $a'$ decreases the region near the origin disappears last.  The region of timelike $S(\tau)$ lies outside the two AHs (larger $R$), though very close to them for quite a range of $r$. 

\begin{figure}[!hb]
\centering 
\includegraphics[width=0.7\textwidth]{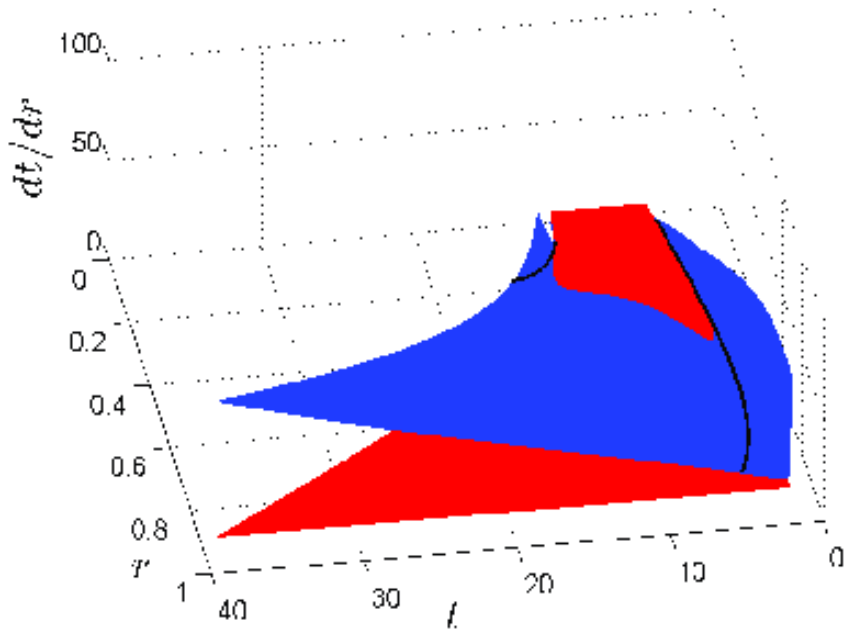}
\caption{Elliptic model (1) with the first coefficient set: The relationship between the $S(\tau)$ (red) and null (blue) surfaces, where their heights represent the slopes of these surfaces.  The plotted time range only includes the crunch at smaller $r$ values.  The edge of the blue surface is irregular due to the clipping of divergence near the bang and crunch.  The black lines represent the past and future apparent horizons.\showlabel{emodel13}}
\end{figure}

To investigate how the size of the timelike region is affected by the crunch time, we next keep the scale time $\tilde{T}$ unchanged, but adjust $a$ to make the crunch time constant.  Thus $M(r)$ and $f(r)$ are as in \er{em11}-\er{em12}, but $a$ becomes
\begin{align}
\showlabel{em43}
a & = -2\pi\tilde{T} = - 2 \pi \frac{M}{f^{3/2}},
\end{align}
The coefficients are changed to $M_{0} = 1$, $M_{1} = 0.1$, $M_{2} = 0.1$, $k = 1$, $f_{1} = -0.67$ and $f_{2} = 0$.  See figures \ref{emodel41} and \ref{emodel42}.

\begin{figure}[!hb]
\centering
\subfloat[The behavior of $M$, $f$, $a$ and $\tilde{T}$\showlabel{emodel41}]{\includegraphics[width=0.5\textwidth]{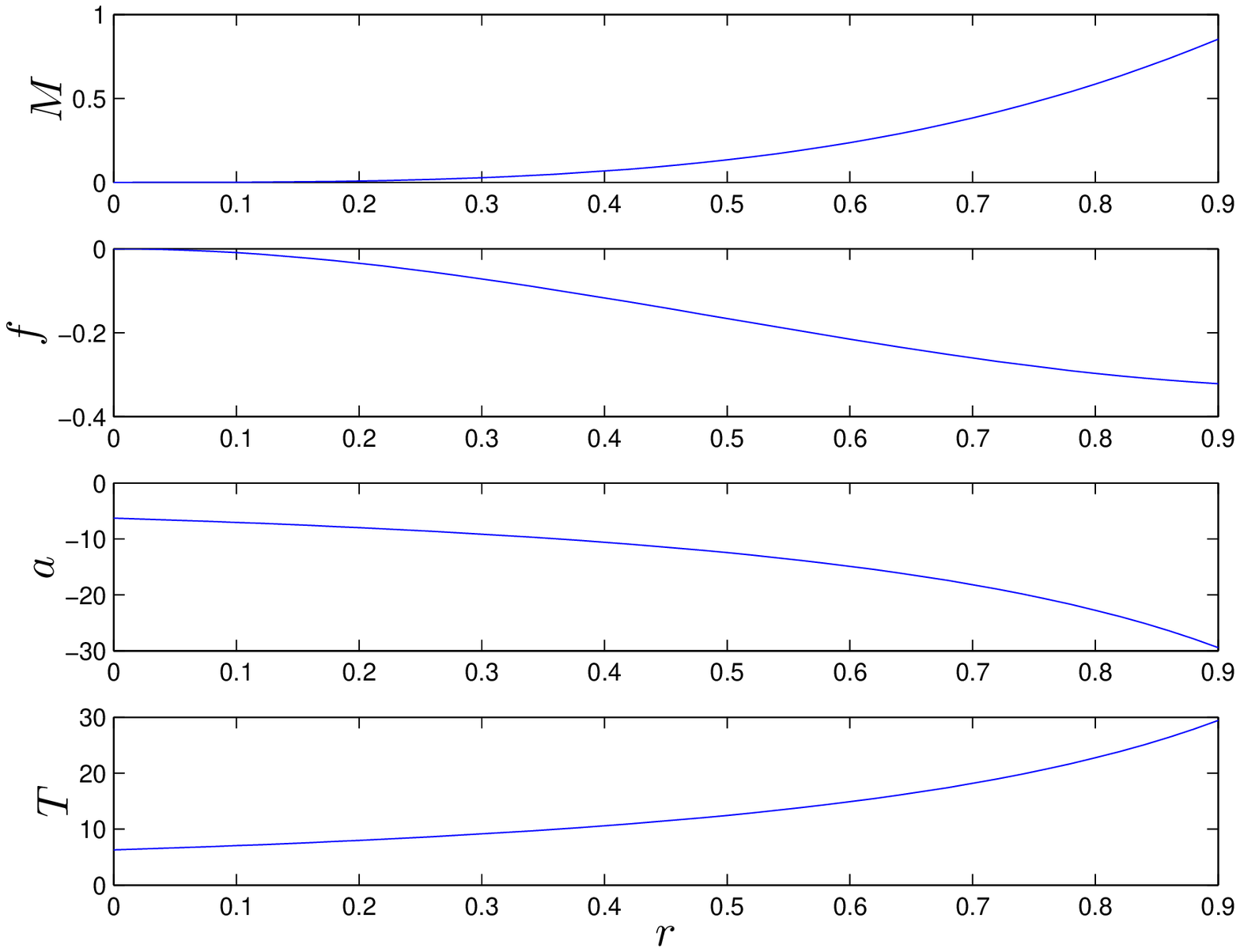}}
\subfloat[Conditions for no shell crossings\showlabel{emodel42}]{\includegraphics[width=0.5\textwidth]{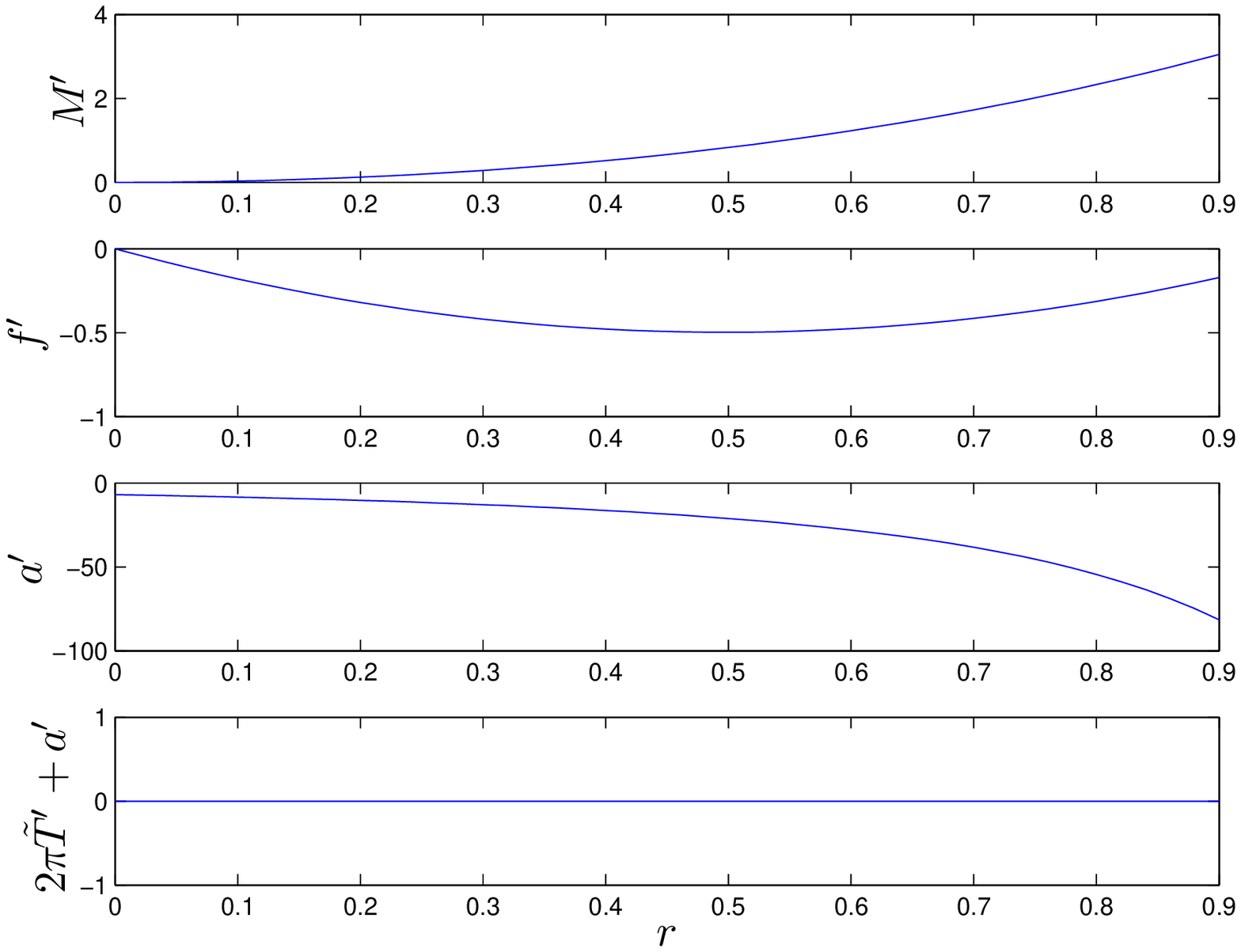}}
\caption{Arbitrary Functions for Elliptic model (1) with the second coefficient set.\showlabel{emodel4}}
\end{figure}

\begin{figure}[!hb]
\centering 
\includegraphics[width=0.7\textwidth]{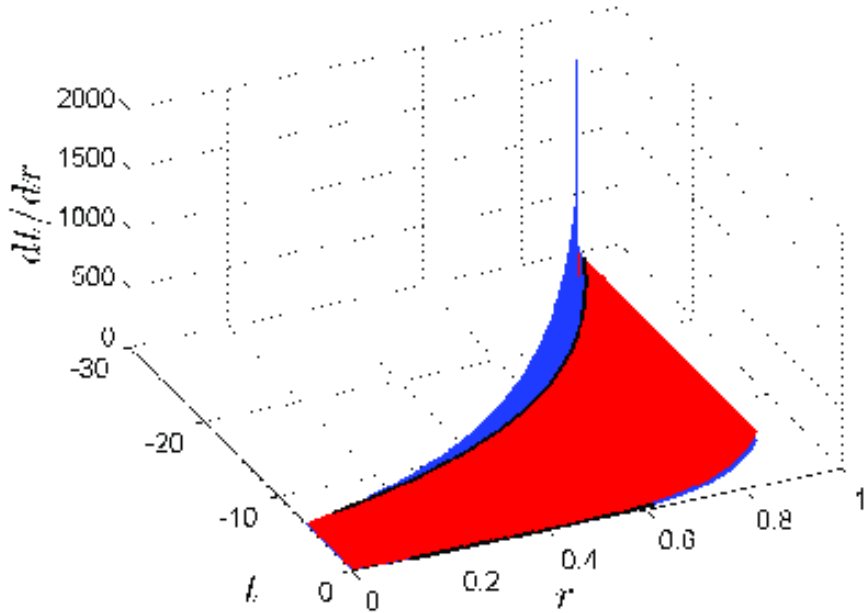}
\caption{Elliptic model (1) with the second coefficient set: The relationship between the $S(\tau)$ and null surfaces.  The crunch is not included in the plotted range for the larger $r$ worldlines.\showlabel{emodel43}}
\end{figure}

Figure \ref{emodel43} shows the $S(\tau)$ surfaces are initially spacelike everywhere, but at intermediate and late times they become timelike, owing to the constant crunch time.

This situation can be changed if $a$ is made a different multiple of lifetime $\tilde{T}$.  For instance, if $a = -0.9(2\pi\tilde{T})$, the $S(\tau)$ are spacelike at early times, and stay timelike through intermediate up to quite late times. 
If $a = -0.2(2\pi\tilde{T})$, the $S(\tau)$ become timelike at intermediate times and return to spacelike quite soon thereafter, as shown in figure \ref{emodel43a}.  In this case, we find two discrete timelike regions; one close to the past AH, the other is close to future AH and the origin.  If $a = -0.01(2\pi\tilde{T})$, the $S(\tau)$ become spacelike everywhere.  As always, the timelike region or regions are always outside the apparent horizons (the black lines), though often very close to them.

\begin{figure}[!hb]
\centering 
\includegraphics[width=0.7\textwidth]{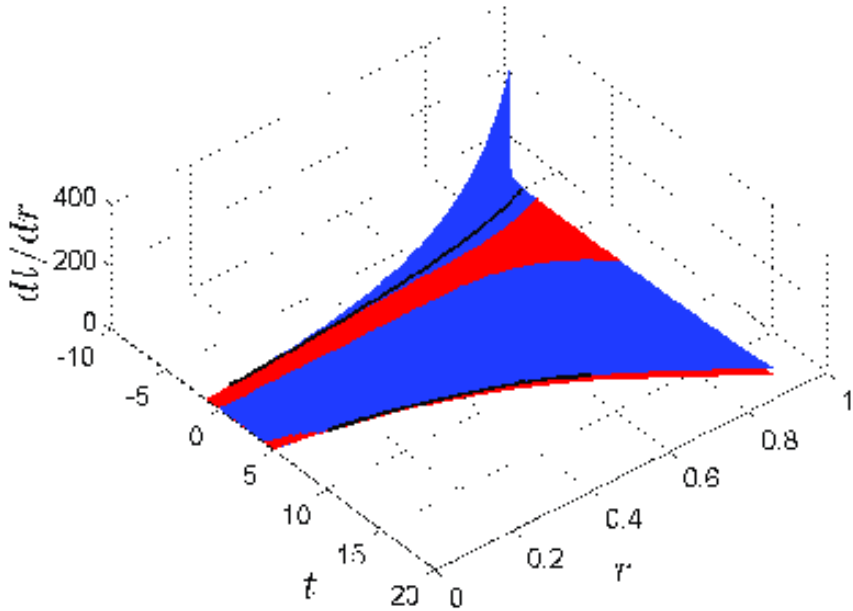}
\caption{Elliptic model (1) with the second coefficient set and $a = -0.2(2\pi\tilde{T})$: Comparison of the $S(\tau)$ and null surfaces, showing the $S(\tau)$ are timelike only at intermediate times. The plotted evolution range does not reach the crunch at larger $r$ values.\showlabel{emodel43a}}
\end{figure}

%Figure \ref{e4} shows the variation of the timelike region size as the strength of the bang time variation, relative to the lifetime, is decreased; $a = -\alpha(2\pi\tilde{T})$ with $\alpha$ varying. The value $\alpha = 1$ gives a constant crunch time.  Generally in this model, the timelike region extends to late times (i.e. the crunch) only if the crunch time is constant. The size of the timelike region can be increased by increasing $(-a')$ and vice versa, but it is confined to be outside the AHs, though often lying very close to them.

%\begin{figure}[!hb]
%\centering
%\subfloat[$a = -0.9(2\pi\tilde{T})$\showlabel{e41}]{\includegraphics[width=0.33\textwidth]{surface_plot_r_elliptic_4a_2.eps}}
%\subfloat[$a = -0.2(2\pi\tilde{T})$\showlabel{e42}]{\includegraphics[width=0.33\textwidth]{surface_plot_r_elliptic_4b_2.eps}}
%\subfloat[$a = -0.01(2\pi\tilde{T})$\showlabel{e43}]{\includegraphics[width=0.33\textwidth]{surface_plot_r_elliptic_4c_2.eps}}
%\caption{Elliptic model (1) with the second coefficient set and different bang times: The extent of the timelike $S(\tau)$ region is strongly affected by this variation.\showlabel{e4}}
%\end{figure}

%%%%%%%%%%%%%%%%%%%%%%%%%%%%%%%%%%%%%%%%%%%%%%%%%%%%%%%%%%%%%%%%%%%%%%%%%%%%%%%%%%%%%%%%%%%%%%%%%%%%
\subsubsection{Model 2}
For this model the arbitrary functions are 
\begin{align}
\showlabel{em21}
M & = M_{0} + M_{1} \exp(r^{2}/M_{2}),\\
\showlabel{em22}
f & = - k \exp(r^{2}/f_{2}),\\
\showlabel{em23}
a & = a_{2}r^{2},
\end{align}
where $M_{0} = 2$, $M_{1} = 1$, $M_{2} = 0.1$, $k = 1$, $f_{2} = 0.1$ and $a_{2} = -1$; see figures \ref{emodel31} and \ref{emodel32}.

\begin{figure}[!hb]
\centering
\subfloat[The behavior of $M$, $f$, $a$ and $\tilde{T}$\showlabel{emodel31}]{\includegraphics[width=0.5\textwidth]{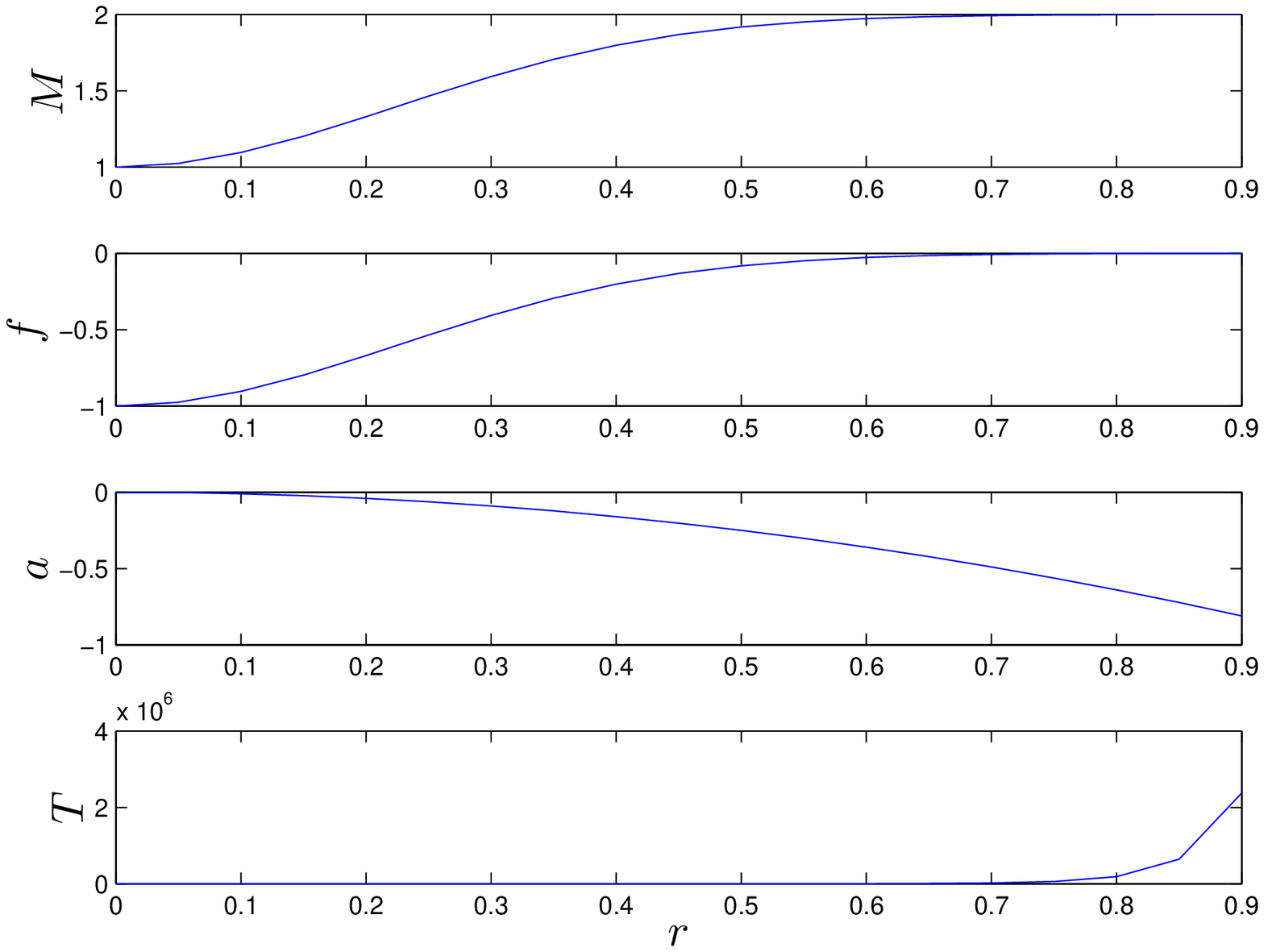}}
\subfloat[The no shell crossing conditions\showlabel{emodel32}]{\includegraphics[width=0.5\textwidth]{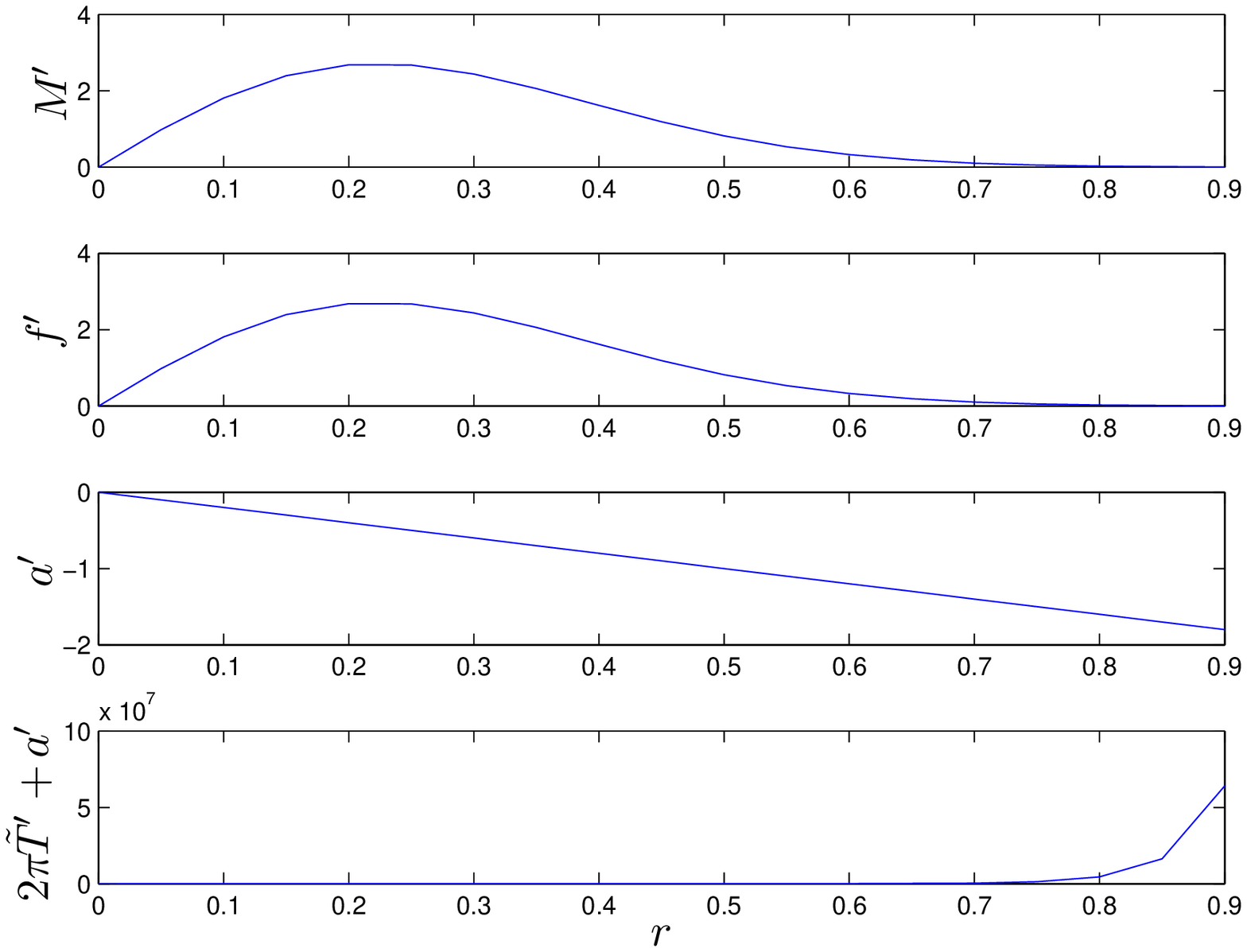}}
\caption{Arbitrary Functions for Elliptic model (2); a non-vacuum black hole.\showlabel{emodel3}}
\end{figure}
This set of functions represents a non-vacuum black hole; it has $f(0) = -1$ and $f'(0) = 0 = M'(0) = a'(0)$, as required to get the full manifold with two asymptotic regions joined by a ``neck"  \cite{Hellaby_1999}.  There are two asymptotic regions, and matter flows out of the past singularity (or bang) and into the future singularity (or crunch).  Figure \ref{emodel33} shows the $S(\tau)$ surfaces become timelike at intermediate times, outside the black hole, while at early and late times the $S(\tau)$ are spacelike everywhere.  A vacuum black hole is obtained by putting $M' = 0$, but note that it is not appropriate for testing Ellis's idea because there are no unique timelike Ricci eigenlines.  

\begin{figure}[!hb]
\centering 
\includegraphics[width=0.7\textwidth]{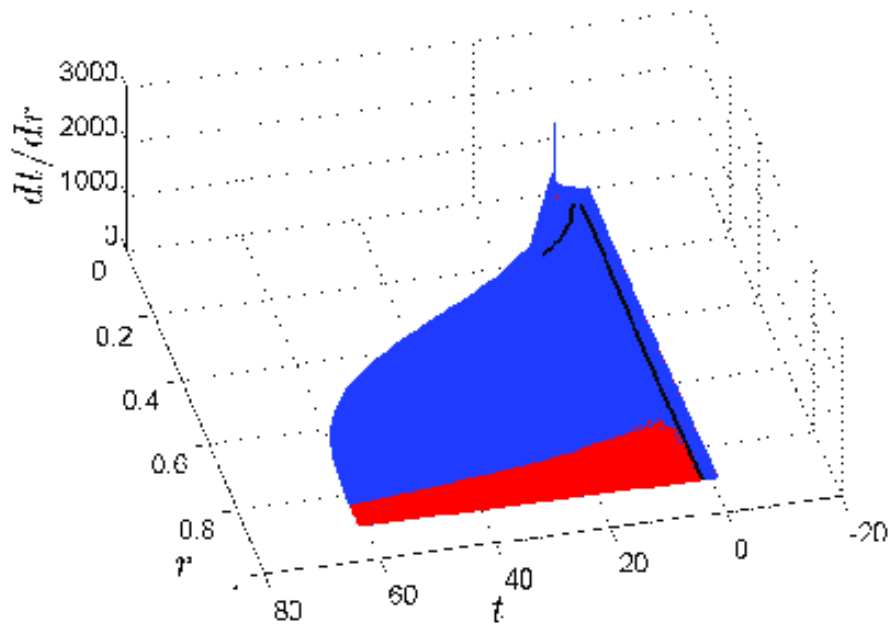}
\caption{Elliptic model (2): The character of the $S(\tau)$ and null surfaces for a non-vacuum black hole.
\showlabel{emodel33}}
\end{figure}

%%%%%%%%%%%%%%%%%%%%%%%%%%%%%%%%%%%%%%%%%%%%%%%%%%%%%%%%%%%%%%%%%%%%%%%%%%%%%%%%%%%%%%%%%%%%%%%%%%%%
\subsection{Discussion}

We have found in all cases that the \{$\tau$= constant\} surfaces $S(\tau)$ are spacelike everywhere at early times.

For the hyperbolic models examined, timelike regions seem common at intermediate times unless $a'$ is very small.  However the $S(\tau_c)$ are nearly always spacelike in the late time limit, except for the rather special case that $M' = f' = 0$ on certain worldlines.  For worldlines that are very close to this condition, the timelike character of the $S(\tau)$ can remain until for a long period.  Now $M' = 0$ ensures zero density locally, but if it's only at one $r$ value, the Ricci eigenline still exists as a limit.

For elliptic cases, intermediate times could easily have timelike $S(\tau)$ regions, but to get timelike $S(\tau)$ at late times (the crunch), the crunch time must be a constant, and the bang time not.  It is notable that with black hole and black-hole-like models, the regions of timelike $S(\tau)$ do not occur along worldlines near to the ``neck" but they do occur further out.

In the models looked at, the timelike regions always occur outside the horizon, but often very close to it.

Invariably, the size of a timelike region was increased by increasing $(-a')$ and vice versa, vanishing completely for samll enough $(-a')$.  Additionally, as $(-a')$ decreases, the timelike region shrinks towards the origin.

%% file: Section5.tex
\section{Conclusion}
\showlabel{Section5}

The true nature of the past, present and future is a mystery, and how the flow of time can best be represented diagrammatically remains problematic, so there has been a great effort by many authors attempting to envisage the flow of time and give a clear picture of our universe \cite{Butterfield:2001fh, FQXi_2014, Mellor98}. According to the block universe idea \cite{Ellis:2012ay}, the universe consists of a unique family of surfaces $S(\tau)$, and the present is merely one surface of constant $\tau$. Conditions at the present time can be evaluated by integration from known conditions at any earlier time, possibly the beginning, up to the present time. The evolving block universe (EBU) and the crystallising block universe (CBU) take the second law of thermodynamics and quantum effects seriously, and replace the time-reversible, deterministic model with one that recognises the past is fixed while the future is highly mutable \cite{Ellis_2006, Ellis_Rothman_2010}.

Our main purpose was to investigate the viability of a particular proposition of Ellis \cite{Ellis:2012ay}, that the correct time $\tau$ to use is the proper time along the timelike eignevectors of the Ricci tensor (``Ricci time").  It was suggested that if the resulting constant $\tau$ surfaces ``sensed" strong gravitational fields, for example by becoming timelike, this would indicate the right kind of behaviour. 

The \LT\ (LT) metric \cite{Hellaby:2009vz, andrzej97, Hellaby_1985} was used to generate a variety of models with different features; some like inhomogeneous cosmologies, and some like matter-filled black holes. The slopes of the constant $\tau$ surfaces $S(\tau)$ were compared with the slopes of the radial null surfaces in order to determine if they were spacelike, null or timelike.

Analytic methods were used to find conditions, on the LT arbitrary functions, $M(r)$, $f(r)$ \& $a(r)$, for the $S(\tau)$ to become timelike in two limits: early times near the bang, and late times near the crunch for elliptic models or $\tau\to\infty$ for hyperbolic models.  Intermediate time calculations, though not definitive, indicated it was rather easy to make $S(\tau)$ timelike.  Numerical methods were then used to calculate and display the full evolution of this relationship for a variety of explicit LT models.

In summary, the $S(\tau)$ surfaces are spacelike everywhere at early times for models that are free of irregularities; it is already known that the bang surface is spacelike.  In the late time limit, the $S(\tau)$ can be timelike, though only under quite strong conditions, and these conditions are totally different in hyperbolic and elliptic models.  In the hyperbolic case, both the geometry/energy function $f(r)$ and the interior mass function $M(r)$ must be (locally) constant, $M' = 0 = f'$. This means there is vacuum $\rho = 0$ on these worldlines.  In the elliptic case, the crunch time must be constant (as it is in FLRW models), while the bang time is not constant.

We found that timelike $S(\tau)$ regions are common at intermediate times. The ``area" of the timelike $S(\tau)$ regions (in the $(t,r)$ plane) strongly depends on $(-a')$. The range and duration of the timelike $S(\tau)$ regions is increased by making $(-a')$ bigger, and if $(-a')$ is sufficiently small the $S(\tau)$ surfaces will be everywhere spacelike. Also, as $(-a')$ is decreased, the timelike regions sometimes tend to shrink towards the origin (if there is one).

We note that the condition, $a' = 0$, i.e. a constant bang time, which guarantees all constant Ricci time surfaces are spacelike throughout a model, is also the condition which ensures there are no decaying modes in LT models \cite{silk:77}. The growing and decaying modes have been invariantly characterised in \cite{Sussman_2013}; see \cite{Plebanski2006} for the definitions. The asymptotics of LT models, represented by covariant scalars along radial geodesics, including those with $a' = 0$, were investigated in \cite{Sussman:2010a}. Of interest here is the relation to gravitational entropy, which was also investigated for LT models in \cite{Sussman:2014a}. They considered the entropy proposals of \cite{Clifton:2013dha, Hosoya:2004nh} and a modification of the latter in \cite{Sussman:2012xc}.  They found that entropy grows if there are anti-correlations between Hubble and density fluctuations. This requires that the decaying modes are sufficiently suppressed, and near the bang they have to be absent for entropy to grow. They also suggest that regions of decreasing entropy and dominant decaying modes may be associated with instability. A related result \cite{Bolejko:2013doa} is that for generic inhomogeneous models, the inhomogeneity initally decreases and then increases, and that entropy decreases while inhomogeneity decreases.

The timelike $S(\tau)$ regions always appeared outside the apparent horizons, i.e. at larger areal radius, though they were often right next to them.

For one specific elliptic model we found there were two discrete timelike $S(\tau)$ regions on certain worldlines, separated by a significant time. One of them was near the past apparent horizon (AH) and the other was very close to the future AH.

In the black-hole-like models, we found that the timelike regions not only stayed outside the AHs, but also did not occur on worldlines at or near the ``neck".

Overall, we found that the Ricci time has the opposite of the suggested behaviour, and if its \{constant $\tau$\} surfaces $S(\tau)$ become timelike, this is in regions outside the apparent horizons, where the gravitational fields are not so strong. Therefore this time may not be appropriate for the EBU.

%% file: RicciTimeInLT.bbl
\begin{thebibliography}{10}

\bibitem{FQXi_2014}
Fqxi forum: The nature of time essay contest.
\newblock (2014).

\bibitem{Alnes_Amarzguioui_Gron_2006}
H\r{a}vard Alnes, Morad Amarzguioui, and {\O}yvind Gr{\o}n.
\newblock Inhomogeneous alternative to dark energy?
\newblock {\em Physical Review D}, 73(8):083519, (2006).

\bibitem{barbour99}
Julian Barbour.
\newblock {\em The End of Time: The Next Revolution in Phyiscs}.
\newblock Oxford University Press, (1999).

\bibitem{Bolejko_Krasinski_Hellaby_2005}
Krzysztof Bolejko, Andrzej Krasi\'{n}ski, and Charles Hellaby.
\newblock Formation of voids in the universe within the lema\^{\i}tre-tolman
  model.
\newblock {\em Monthly Notices of the Royal Astronomical Society},
  362:213--228, (2005).

\bibitem{Bolejko:2013doa}
Krzysztof Bolejko and William~R. Stoeger.
\newblock Intermediate homogenization of the universe and the problem of
  gravitational entropy.
\newblock {\em Physical Review D}, 88:063529, (2013).

\bibitem{Bondi_1947}
Hermann Bondi.
\newblock Spherically symmetrical models in general relativity.
\newblock {\em Monthly Notices of the Royal Astronomical Society}, 107:410,
  (1947).

\bibitem{Bonnor1985}
William~B. Bonnor.
\newblock Closed tolman models of the universe.
\newblock {\em Classical and Quantum Gravity}, 2:781--790, (1985).

\bibitem{Butterfield:2001fh}
Jeremy~N. Butterfield.
\newblock The end of time?
\newblock (2001).

\bibitem{Christodoulou_1984}
Demetrios Christodoulou.
\newblock Violation of cosmic censorship in the gravitational collapse of a
  dust cloud.
\newblock {\em Communications in Mathematical Physics}, 93:171--95, (1984).

\bibitem{Clifton:2013dha}
Timothy Clifton, George~F.R. Ellis, and Reza Tavakol.
\newblock A gravitational entropy proposal.
\newblock {\em Classical and Quantum Gravity}, 30:125009, (2013).

\bibitem{Davies_2012}
Paul C.~W. Davies.
\newblock That mysterious flow.
\newblock {\em Scientific American Special Edition: A Matter of Time},
  21:8--13, (2012).

\bibitem{Eardley_1979}
Douglas~M. Eardley and Larry Smarr.
\newblock Time functions in numerical relativity: Marginally bound dust
  collapse.
\newblock {\em Physical Review D}, 19:2239--59, (1979).

\bibitem{Ellis_2006}
George F.~R. Ellis.
\newblock Physics in the real universe: Time and spacetime.
\newblock {\em General Relativity and Gravitation}, 38(12):1797--1824, (2006).

\bibitem{Ellis:2012ay}
George F.~R. Ellis.
\newblock Space time and the passage of time.
\newblock (2012).

\bibitem{Ellis_Rothman_2010}
George F.~R. Ellis and Tony Rothman.
\newblock Time and spacetime: The crystallizing block universe.
\newblock {\em International Journal of Theoretical Physics}, 49(5):988--1003,
  (2010).

\bibitem{Ellis_1967}
George~F.R. Ellis.
\newblock Dynamics of pressure-free matter in general relativity.
\newblock {\em Journal of Mathematical Physics}, 8:1171, (1967).

\bibitem{Hellaby_1985}
Charles Hellaby.
\newblock Some properties of singularities in the tolman model.
\newblock {\em Ph.D. thesis, Queen’s University at Kinston, Ontario}, (1985).

\bibitem{Hellaby_1999}
Charles Hellaby.
\newblock A kruskal-like model with finite density.
\newblock {\em Classical and Quantum Gravity}, 4:635--650, (1987).

\bibitem{Hellaby_1994}
Charles Hellaby.
\newblock On the vaidya limit of the tolman model.
\newblock {\em Physical Review D}, 49:6484--6488, (1994).

\bibitem{Hellaby_1996}
Charles Hellaby.
\newblock The nonsimultaneous nature of the schwarzschild r = 0 singularity.
\newblock {\em Journal of Mathematical Physics}, 37:2892, (1996).

\bibitem{Hellaby_2006m}
Charles Hellaby.
\newblock The mass of the cosmos.
\newblock {\em Monthly Notices of the Royal Astronomical Society},
  370:239--244, (2006).

\bibitem{Hellaby:2009vz}
Charles Hellaby.
\newblock Modelling inhomogeneity in the universe.
\newblock {\em Proceedings of Science}, ISFTG:005, (2009).

\bibitem{Hellaby_Alfedeel_2009}
Charles Hellaby and Alnadhief H.~A. Alfedeel.
\newblock Solving the observer metric.
\newblock {\em Physical Review D}, 79:043501, (2009).

\bibitem{Hellaby_Lake_1986}
Charles Hellaby and Kayll Lake.
\newblock Shell crossings and the tolman model.
\newblock {\em Astrophysical Journal}, 290:381--387, (1985).

\bibitem{Hosoya:2004nh}
Akio Hosoya, Thomas Buchert, and Masaaki Morita.
\newblock Information entropy in cosmology.
\newblock {\em Physical Review Letters}, 92:141302, (2004).

\bibitem{Joshi93}
Pankaj~S. Joshi.
\newblock {\em Global Aspects in Gravitation and Cosmology}.
\newblock Clarendon Press, Oxford, (1993).

\bibitem{andrzej97}
Andrzej Krasi\'{n}ski.
\newblock {\em Inhomogeneous Cosmological Models}.
\newblock Cambridge University Press, (1997).

\bibitem{Hellaby_Krasinski_2001}
Andrzej Krasi\'{n}ski and Charles Hellaby.
\newblock Structure formation in the lema\^{\i}tre-tolman model.
\newblock {\em Physical Review D}, 65:023501, (2001).

\bibitem{Krasinski_Hellaby_2004a}
Andrzej Krasi\'{n}ski and Charles Hellaby.
\newblock Formation of a galaxy with a central black hole in the lema\^{\i}tre-
  tolman model.
\newblock {\em Physical Review D}, 69:043502, (2004).

\bibitem{Krasinski_Hellaby_2004}
Andrzej Krasi\'{n}ski and Charles Hellaby.
\newblock More examples of structure formation in the lema\^{\i}tre-tolman
  model.
\newblock {\em Physical Review D}, 69:023502, (2004).

\bibitem{Lemaitre_1933}
Georges Lema\^{\i}tre.
\newblock L'univers en expansion.
\newblock {\em Annales de la Soci\'{e}t\'{e} Scientifique de Bruxelles},
  A53:51--85, (1933).

\bibitem{Lemos_1992}
Jos\'{e} P.~S. Lemos.
\newblock Naked singularities: Gravitationally collapsing configurations of
  dust or radiation in spherical symmetry, a unified treatment.
\newblock {\em Physical Review Letters}, 68:1447--50, (1992).

\bibitem{Mellor98}
David~H. Mellor.
\newblock {\em Real Time II}.
\newblock London: Routledge, (1998).

\bibitem{Newman_1986}
Richard P. A.~C. Newman.
\newblock Strengths of naked singularities in tolman-bondi spacetimes.
\newblock {\em Classical and Quantum Gravity}, 3:527, (1986).

\bibitem{Plebanski2006}
J.~Plebanski and A.~Krasi\'{n}ski.
\newblock {\em An Introduction to General Relativity and Cosmology}.
\newblock Cambridge University Press, (2006).

\bibitem{price96}
Huw Price.
\newblock {\em Time’s Arrow and Archimedes’ Point}.
\newblock Oxford University Press, (1997).

\bibitem{Rovelli:2009ee}
Carlo Rovelli.
\newblock Forget time.
\newblock {\em Foundations of Physics}, 41(9):1475--1490, (2011).

\bibitem{silk:77}
Joseph Silk.
\newblock Large-scale inhomogeneity of the universe: Spherically symmetric
  models.
\newblock {\em Astronomy and Astrophysics}, 59:53, (1977).

\bibitem{Sussman_2007}
Roberto~A. Sussman.
\newblock A dynamical system approach to inhomogeneous dust solution.
\newblock {\em Classical and Quantum Gravity}, 25:015012, (2007).

\bibitem{Sussman:2010a}
Roberto~A. Sussman.
\newblock Radial asymptotics of lema\^{\i}tre-tolman-bondi dust models.
\newblock {\em General Relativity and Gravitation}, 42:2813--2864, (2010).

\bibitem{Sussman_2013}
Roberto~A. Sussman.
\newblock Invariant characterisation of the growing and decaying density modes
  in ltb dust models.
\newblock {\em Classical and Quantum Gravity}, 30:235001, (2013).

\bibitem{Sussman:2012xc}
Roberto~A. Sussman.
\newblock Weighed scalar averaging in ltb dust models, part i:statistical
  fluctuations and gravitational entropy.
\newblock {\em Classical and Quantum Gravity}, 30:065015, (2013).

\bibitem{Sussman:2014a}
Roberto~A. Sussman and Julien Larena.
\newblock Gravitational entropies in {LTB} dust models.
\newblock {\em Classical and Quantum Gravity}, 31:075021, (2014).

\bibitem{Tolman_1934}
Richard~C. Tolman.
\newblock Effect of inhomogeneity on cosmological models.
\newblock {\em National Academy of Sciences of the USA}, 20:169, (1934).

\bibitem{Waugh_Lake_1989}
Brad Waugh and Kayll Lake.
\newblock Shell-focusing singularities in spherically symmetric self-similar
  spacetimes.
\newblock {\em Physical Review D}, 40:2137--2139, (1989).

\end{thebibliography}
